\documentclass[12pt, a4paper]{article}
\pdfoutput=1
\usepackage{graphicx}
\usepackage{color}
\usepackage{amssymb}
\usepackage{amsmath}
\usepackage{bm}
\usepackage{comment}
\usepackage{color}
\usepackage[dvipsnames]{xcolor}
\usepackage{slashed}
\usepackage{subfigure}
\usepackage{epstopdf}            
\usepackage{epsfig}
\renewcommand{\thefootnote}{\fnsymbol{footnote}}
\newcommand{\beq}{\begin{equation}}
\newcommand{\eeq}{\end{equation}}

\newcommand{\ol}{\overline}

\newcommand{\eps}{\epsilon}

\newcommand{\Ox}{{\rm O}}
\newcommand{\mutau}{L_\mu-L_\tau}
\newcommand{\TS}{{\rm TS}}

\newcommand{\Slash}[1]{{\ooalign{\hfil/\hfil\crcr$#1$}}}

\newcommand{\VEV}[1]{{\langle #1 \rangle}}

\definecolor{azur}{rgb}{0.118,0.498,0.796}
\definecolor{darkred}{cmyk}{0,1,1,0.4}
\definecolor{green1}{rgb}{0.21,0.6,0.32}

\usepackage[colorlinks=true, linkcolor=magenta, citecolor=ForestGreen, urlcolor=black]{hyperref} 

\begin{document}

\begin{titlepage}

\begin{flushright}
KYUSHU-HET-217
\end{flushright}

\begin{center}

{\Large
{\bf 
Search for U(1)$_{\mutau}$ charged Dark Matter\\ \vspace{1ex} with neutrino telescope}
}

\vskip 2cm

Kento Asai$^{1}$\footnote{
\href{mailto:asai@hep-th.phys.s.u-tokyo.ac.jp}{\tt
 asai@hep-th.phys.s.u-tokyo.ac.jp}},
Shohei Okawa$^{2}$\footnote{
\href{mailto:okawa@uvic.ca}{\tt
 okawa@uvic.ca}}
and
Koji Tsumura$^{3}$\footnote{
\href{mailto:tsumura.koji@phys.kyushu-u.ac.jp}{\tt
 tsumura.koji@phys.kyushu-u.ac.jp}}

\vskip 0.5cm

{\it $^1$ Department of Physics, University of Tokyo, Bunkyo-ku, Tokyo 113--0033, Japan} \\[3pt]

{\it $^2$ Department of Physics and Astronomy, University of Victoria, \\
Victoria, BC V8P 5C2, Canada} \\[3pt]

{\it $^3$
Department of Physics, Kyushu University, 744 Motooka, Nishi-ku, Fukuoka, 819--0395, Japan} 

\vskip 1.5cm

\begin{abstract}
We study a simple Dirac fermion dark matter model in U(1)$_{\mutau}$ theory. 
The new light gauge boson $X$ plays important roles in both dark matter physics 
and the explanation for the muon $g-2$ anomaly. 
The observed dark matter relic density is realized by a large U(1)$_{\mutau}$ charge 
without introducing a resonance effect of the $X$ boson. 
As a by-product of the model, characteristic neutrino signatures from sub-GeV dark matter $\psi$
are predicted depending on the mass spectrum. 
We formulate the analysis of $\psi\bar\psi\to \nu\bar\nu$, and of  
$\psi\bar\psi\to XX$ followed by $X\to\nu\bar\nu$, in a model independent way. 
The energy spectrum of neutrinos in the former process is monochromatic while 
in the latter process is bowl-shape. 
We also evaluate sensitivity at Super-Kamiokande and future Hyper-Kamiokande detectors. 
The analysis is finally applied to the U(1)$_{\mutau}$ dark matter model. 
\end{abstract}

\end{center}
\end{titlepage}

\def\rem#1{ {\bf\textcolor{red}{($\clubsuit$ #1 $\clubsuit$)}}}
\renewcommand{\thefootnote}{\#\arabic{footnote}}
\setcounter{footnote}{0}

\section{Introduction}

High energy collider experiments have confirmed the Standard Model (SM) in particle physics 
with high precision up to the TeV scale. 
In spite of the great success, there are some compelling empirical evidence for new physics, 
one of the most important ones being dark matter (DM). 
In the last decades, 
significant attention has been paid to Weakly Interacting Massive Particles (WIMPs) DM, 
with considerable efforts to directly and indirectly test such DM candidates. 
It is unfortunate, however, that 
all of these experimental efforts have obtained no affirmative signals thus far. 
The resulting stringent constraints often put the pressure on the traditional WIMP DM with the electroweak scale mass. 
This tendency encourages theorists to embark on an unexplored sub-GeV mass region.
A simple realization of the sub-GeV DM includes a secluded scenario~\cite{Pospelov:2007mp} in which 
DM couples to the visible sector only via new light mediator particles, 
such as dark gauge bosons or dark Higgs bosons~\cite{Pospelov:2007mp,Boehm:2003hm,Fayet:2004bw,ArkaniHamed:2008qn,Foot:2014uba,Alves:2016cqf,Escudero:2017yia,Darme:2017glc,Dutra:2018gmv}. 

Another attractive hint for new physics is the so-called muon $g-2$ anomaly. 
There is over $3\,\sigma$ discrepancy between the theoretical prediction and experimental value 
of the muon anomalous magnetic moment in light of the result measured at the BNL E821~\cite{Bennett:2006fi,Roberts:2010cj}.  
This tension suggests that there exist new interactions in the muon sector, 
leading to growing interests on lepton flavored new physics scenarios. 
One well-studied class of extensions is gauged U(1)$_{\mutau}$ models~\cite{Foot:1990mn,He:1990pn,He:1991qd,Foot:1994vd}. 
The associated gauge boson with the sub-GeV mass 
has a sizable contribution to accommodate the discrepancy, 
while avoiding existing constraints 
from collider experiments and cosmological and astrophysical observations because of the absence of direct coupling with electron and quarks~\cite{Baek:2001kca,Ma:2001md,Heeck:2011wj,Harigaya:2013twa}. 
Next generation low-energy experiments will have the sensitivity to most of the parameter space 
that addresses the discrepancy. 

Given the potential of the muon $g-2$ explanation and 
the increased interest in the secluded DM scenario, 
it is natural to ask if the U(1)$_{\mutau}$ gauge boson can be a mediator between DM and the SM sector. 
In the present paper, we therefore consider a SM singlet vectorlike fermion DM $\psi$, 
charged under the U(1)$_{\mutau}$ group with the associated gauge boson $X$, 
and scrutinize the phenomenology. 
Because DM is introduced vectorlike, gauge anomalies relative to U(1)$_{\mutau}$ group, which come from DM contributions, are cancelled automatically.
Similar DM candidates have extensively been studied in the literature, 
mostly in the traditional (heavy) WIMP regime~\cite{Baek:2008nz,Altmannshofer:2016jzy,Arcadi:2018tly,Bauer:2018egk} and 
by introducing additional new degrees of freedom for the successful DM production~\cite{Baek:2015fea,Patra:2016shz,Biswas:2016yan}. 
The light mass window is studied in \cite{Foldenauer:2018zrz}, 
under the assumption with no charge hierarchy between DM and muon. 
It is shown that the resonant DM annihilation via the $X$ boson physical pole is required to 
correctly produce the DM relic density in the muon $g-2$ favored region. 
In contrast to these works, we do not restrict the DM gauge charge to unity and 
not introduce any additional fields except for DM and the gauge boson. 
We then scan the whole parameter space. 
As a result, 
we find that allowing a charge hierarchy opens a new possibility 
that both the muon $g-2$ and DM relic abundance are explained
without the resonance enhancement. 
It also turns out that the DM mass is restricted to be 10\,MeV--1\,GeV, 
in light of the direct detection experiments and the Cosmic Microwave Background (CMB) observations. 

We also examine the indirect detection signals of the DM candidate at the Super-Kamiokande (SK) and Hyper-Kamiokande (HK). 
It is remarkable that the model can be tested with neutrino telescopes 
by searching for an extra neutrino flux from galactic DM annihilation. 
The main annihilation mode of DM is the one into neutrinos or $X$ bosons, 
depending on the mass spectrum of DM and $X$. 
In the former annihilation, the produced neutrino flux is monochromatic at $E_\nu=m_\psi$, 
providing a sharp DM signal. 
In the latter one, the neutrinos are produced via the 1-step cascade process, 
$\psi\bar{\psi} \to XX \to 2\nu 2\bar{\nu}$. 
The shape of the produced neutrino flux depends on the polarizations of the $X$ boson and takes the bowl-like form as a function of neutrino energy~\cite{Garcia-Cely:2016pse}. 
In the present paper, we analyze both the monochromatic and bowl-like neutrino spectra, and 
derive upper limits on the annihilation cross sections by 
reinterpreting the results of the supernova relic neutrino (SRN) searches at the SK~\cite{Malek:2002ns,Bays:2011si}. 
We would like to emphasize that 
while there are many similar studies with a special focus on the monochromatic spectrum~\cite{Aartsen:2016pfc,Aartsen:2017ulx,Baur:2019jwm,Abe:2020sbr,ANTARES:2019svn,Yuksel:2007ac,PalomaresRuiz:2007eu,Primulando:2017kxf,Campo:2017nwh,Campo:2018dfh,Klop:2018ltd,Arguelles:2019ouk,Bell:2020rkw}, 
the analysis for the bowl neutrino spectrum is given for the first time in this paper. 
Although we analyze the neutrino flux originated from DM annihilation under the U(1)$_{\mutau}$ DM model in the present paper, such analyses can be applied to non-flavored U(1) group and other kinds of DM.
We further study the future sensitivity of the HK experiment and 
discuss the impacts on the U(1)$_{\mutau}$ DM model. 
It is possible that the HK with an updated option of the doped Gd will 
probe the canonical thermal relic cross section in the ${\cal O}$(10)\,MeV region. 

This paper is organized as follows. 
In Sec.~\ref{sec:model}, we introduce the model Lagrangian involving two new particles, 
a Dirac DM and a massive gauge boson. 
The constraints on the new gauge boson are also discussed there. 
We study the DM phenomenology in Sec.~\ref{sec:DM}, and 
examine the indirect detection signals of the DM annihilation at the SK and HK experiments in Sec.~\ref{sec:indirect}. 
We summarize our results in Sec.~\ref{sec:summary}.

\section{Model}
\label{sec:model}

The model is built on an SU(3)$_c\times$SU(2)$_L \times$U(1)$_Y\times$U(1)$_{\mutau}$ gauge theory.
We introduce a SM singlet Dirac DM $\psi$ that carries an $\mutau$ charge $q_\psi$ 
in unit of the muon (tau) charge being $+1\,(-1)$. 
The charge assignment for DM and leptons is summarized in Table \ref{table1}. 
The other SM fields have no $\mutau$ charge. 
The DM stability is guaranteed by an accidental global U(1) symmetry. 
In this paper, we assume that neutrinos can approximate massless because their masses are much lighter than the other particles.
In fact, neutrino oscillation has been observed by various experiments, and then neutrinos have undoubtedly non-zero masses. 
There are some previous works which discuss neutrino masses in the U(1)$_{\mutau}$ model, for example Ref.~\cite{Choubey:2004hn,Araki:2012ip,Heeck:2014sna,Crivellin:2015lwa,Asai:2017ryy,Asai:2018ocx,Asai:2019ciz}.
However, neutrino masses are irrelevant in the following discussion about DM, and thus we neglect them.

The model Lagrangian is given by 
\begin{align}
{\cal L} & = {\cal L}_{\rm SM} 
- g_X X_\lambda \left( \ol{\mu} \gamma^\lambda \mu - \ol{\tau} \gamma^\lambda \tau 
+ \ol{\nu_{\mu L}} \gamma^\lambda \nu_{\mu L} - \ol{\nu_{\tau L}} \gamma^\lambda \nu_{\tau L} \right) \nonumber\\
& \quad - \frac{1}{4} X_{\mu\nu} X^{\mu\nu} + \frac{1}{2} m_X^2 X_\mu X^\mu 
- \frac{\eps}{2} X_{\mu\nu} B^{\mu\nu} \nonumber\\
& \quad + \ol{\psi} \left( i \Slash{\partial} - m_\psi \right) \psi - q_\psi g_X X_\lambda \ol{\psi} \gamma^\lambda \psi ,
\end{align}
where $X^\mu$ denotes the U(1)$_{L_\mu-L_\tau}$ gauge boson with the field strength $X^{\mu\nu}$. 
The mass of $X$ is generated by the Higgs mechanism or St\"{u}ckelberg mechanism. 
Since the origin of the mass is irrelevant to our discussion,  
we do not specify it. 
Now, we have five free parameters, $m_X$, $g_X$, $m_\psi$, $q_\psi$ and $\epsilon$ in the model. 
In this paper, we will consider $q_\psi$ as a free parameter that takes an arbitrary value unless it violates unitarity. 
As will see in Sec.~\ref{sec:DM}, a large $q_\psi$ is indeed required to explain the observed DM abundance. 
Such a large $q_\psi$ is allowed phenomenologically, but it may be unnatural from the theoretical perspective. 
We do not discuss in this paper any UV origin of the large charge hierarchy between DM and muon, but 
one of possible UV extensions includes the addition of another dark U(1) gauge symmetry that is coupled only to DM. 
If DM is not charged under the U(1)$_{\mutau}$ symmetry and there is a small mixing between two U(1)'s, we will be able to apparently realize the large charge hierarchy.

\begin{table}[t]
\centering 
\begin{tabular}{c|ccc}\hline
fields & ~~SU(2)$_L$~~  & ~~U(1)$_Y$~~ & ~~U(1)$_{\mutau}$~~ \\ \hline\hline 
   $(L_e, L_\mu, L_\tau)$   &    ${\bf2}$      &         $-1/2$  &       $(0,+1,-1)$           \\
    $(e_R, \mu_R, \tau_R)$   &    ${\bf1}$      &         $-1/2$  &       $(0,+1,-1)$            \\
     $\psi$    &${\bf 1}$      &         $0$      &         $q_\psi$              \\
    \hline 
\end{tabular}
\caption{Charge assignment}
\label{table1} 
\end{table}

We further assume the kinetic mixing $\eps$ of $X$ and the hypercharge gauge field $B$ is vanishing at some high scale. 
Nonetheless, finite mixings of $X$ with photon $A$ and with the $Z$ boson arise at the one-loop level at low energy:
\begin{equation}
{\cal L}_\eps = - \frac{\eps_A}{2} F_{\mu\nu} X^{\mu\nu} - \frac{\eps_Z}{2} Z_{\mu\nu} X^{\mu\nu} ,
\end{equation}
where $F_{\mu\nu}$ and $Z_{\mu\nu}$ denote the field strength for photon and $Z$ boson, respectively. 
We will take into account this one-loop level mixing in our analysis. 
The mixing with photon is calculated below the muon mass scale as 
\begin{equation}
\epsilon_A = - \frac{eg_X}{12\pi^2} \ln\left(\frac{m_\tau^2}{m_\mu^2}\right) .
\end{equation}
To get the canonically normalized gauge fields, 
we shift the photon field as $A_\mu \to A_\mu + \epsilon_A X_\mu$ for $|\eps_A|\ll1$. 
This shift induces the interaction of $X$ to the electromagnetic current $\epsilon_A e X_\mu J_{\rm em}^\mu$, 
which is crucial for DM direct detection. 
Similarly, the mixing with the $Z$ boson is given by 
\begin{equation}
\epsilon_Z = - \left( -\frac{1}{4} + \sin^2\theta_W \right) \frac{eg_X}{12\pi^2  \cos\theta_W\sin\theta_W} \ln\left(\frac{m_\tau^2}{m_\mu^2}\right) ,
\label{eq;epsZ}
\end{equation}
with the Weinberg angle $\theta_W$. 
Shifting $Z_\mu \to Z_\mu - \eps_Z m_X^2/m_Z^2 X_\mu$ and $X_\mu \to X_\mu + \eps_Z Z_\mu$ 
to get the canonically normalized gauge fields, 
we find the $Z$ boson couples to $\mutau$ current and the $X$ boson to the neutral current, 
\begin{align}
{\cal L}_{Z,\mutau} & = - g_X \eps_Z Z_\mu J^\mu_{\mutau} ,\\
{\cal L}_{X,{\rm NC}} &= \frac{g}{\cos\theta_W} \eps_Z \frac{m_X^2}{m_Z^2} X_\mu J^\mu_{\rm NC} .
\end{align}
Since the effect of the mixing $\eps_Z$ is suppressed by $m_X^2/m_Z^2$ and is negligible in the phenomenological study below, we ignore it in this paper. 

\begin{figure}[t]
\centering
\includegraphics[width=0.55\textwidth]{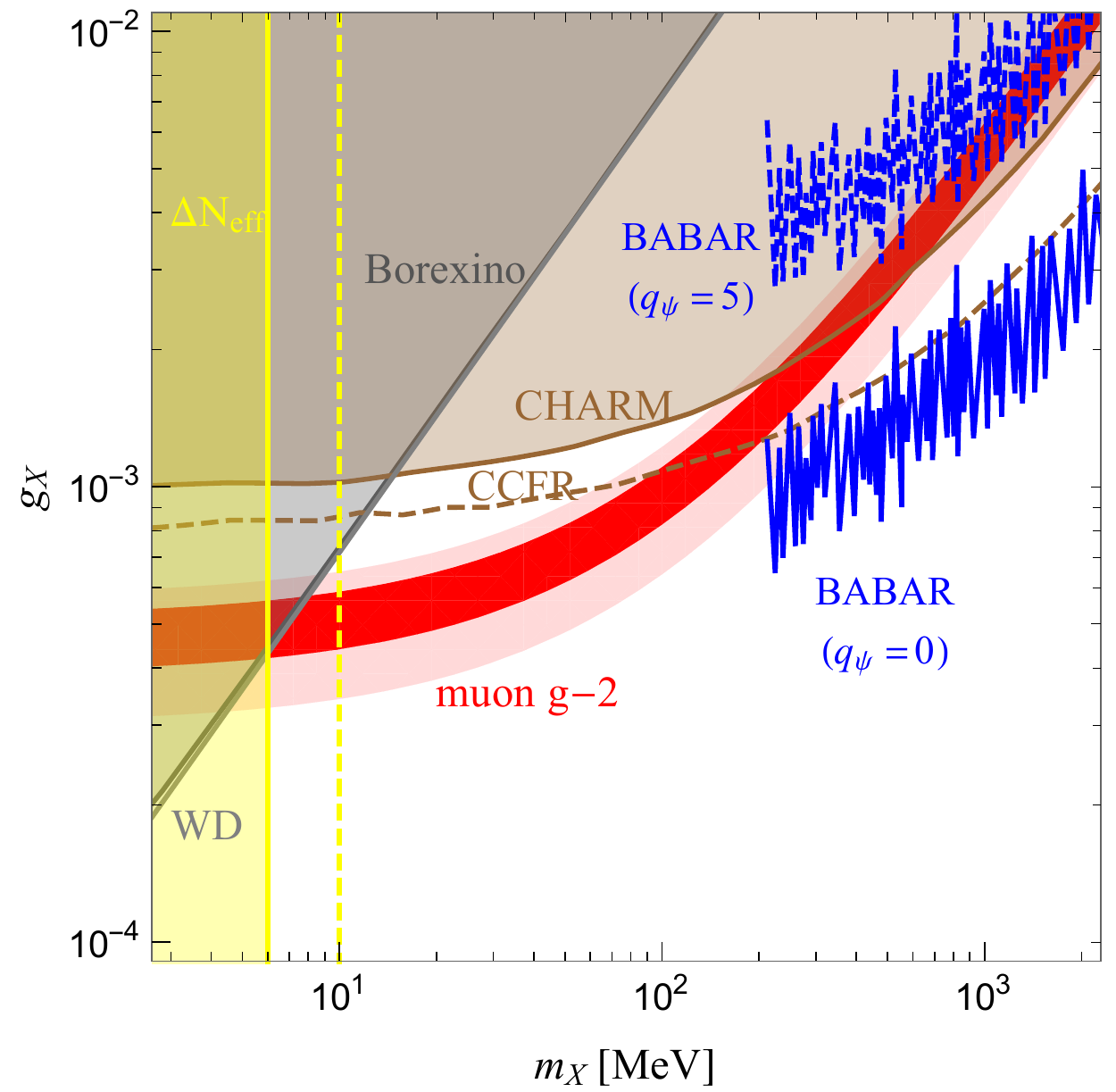}
\caption{
Constraints on the $X$ boson and muon $g-2$ favored region.
Dark (light) red region is favored the muon $g-2$ at $1\sigma \, (2\sigma)$ level.
The other colored regions and lines show regions excluded by the various experimental and astrophysical bounds.
}
\label{fig;Xboson}
\end{figure}

Let us next discuss impacts of the $X$ boson on the muon $g-2$ anomaly 
and other experimental results. 
It can contribute to the anomalous magnetic moment of muon $a_\mu = (g_\mu-2)/2$. 
The one-loop contribution is evaluated as 
\begin{equation}
\Delta a_\mu^X = \frac{g_X^2}{8\pi^2} \int_0^1 dx \frac{2m_\mu^2 x^2(1-x)}{x^2 m_\mu^2 + (1-x) m_X^2} .
\end{equation}
The latest result announced by E821~\cite{Bennett:2006fi,Roberts:2010cj} and 
the theoretical calculation find a 3.3$\sigma$ discrepancy~\cite{Zyla:2020zbs}, 
\begin{equation}
\Delta a_\mu = a_\mu^{\rm exp} - a_\mu^{\rm SM} = (261\pm79)\times10^{-11} .
\end{equation} 
Since $\Delta a_\mu^X \simeq g_X^2/(8\pi^2)$ for $m_X \ll m_\mu$, 
the discrepancy is resolved with $g_X \simeq 5\times10^{-4}$. 
Thus, a finite but small $g_X$ is favored with respect to the muon $g-2$. 
In Fig.~\ref{fig;Xboson}, we show the parameter region, 
in which the muon $g-2$ is explained within $1\sigma$ ($2\sigma$), with the (light) red band. 
We also illustrate the experimental upper limits on $g_X$ from 
a BABAR search for $e^+e^- \to \mu\bar{\mu} X$ with a subsequent decay $X\to \mu\bar{\mu}$~\cite{TheBABAR:2016rlg}, 
CHARM-II and CCFR measurements of the neutrino trident production $\nu N \to \nu N \mu\bar{\mu}$~\cite{Geiregat:1990gz,Mishra:1991bv,Altmannshofer:2014pba}, 
a Borexino measurement of the interaction rate of the the mono-energetic 862\,keV {$^7$Be} solar neutrino~\cite{Bellini:2011rx,Harnik:2012ni,Kaneta:2016uyt}, and the white dwarf (WD) cooling induced by the plasmon decay via the off-shell $X$ boson~\cite{Dreiner:2013tja,Bauer:2018onh}.
Note that the BABAR limit depends on the branching fraction of the $X\to \psi\bar{\psi}$ decay, 
because the BABAR experiment searches for a dark photon decaying into a muon pair. 
As the fraction of the $X\to \psi\bar{\psi}$ decay is larger, the limit is weaker. 
We show in the figure the BABAR limit for $q_\psi=5$ with the dashed blue line, 
assuming DM is much lighter than $X$. 
The neutrino trident, Borexino and WD bounds are not influenced by the existence of DM, 
since the off-shell DM can only contribute to them. 
Hereafter, we focus on the parameter space where the muon $g-2$ is explained 
within $2\sigma$ while avoiding the other constraints 
from the BABAR, CHARM, Borexino experiments and WD cooling. 

There is a cosmological bound on the $X$ boson in addition to the above experimental bounds. 
The $X$ boson mainly decays into a pair of neutrinos for $m_X < 2m_\psi$. 
The lifetime is much shorter than the time scale ($\tau_{\rm BBN} \sim 1$\,sec) 
of Big Bang Nucleosynthesis (BBN) 
in the parameter space that we are interested in. 
If the $X$ boson is light enough, 
the $X$ boson can be in equilibrium with the neutrinos after the neutrino decoupling ($T\sim1$\,MeV). 
Then, it is possible to increase the effective neutrino number $N_{\rm eff}$. 
Since the $X$ boson is already non-relativistic at the BBN, 
it does not contribute directly to $N_{\rm eff}$. 
It can decay into neutrinos, however. 
The decay releases the energy into neutrinos, reheating the neutrino temperature. 
Following \cite{Kamada:2018zxi}, we estimate the contribution to $N_{\rm eff}$ 
and obtain the lower mass bound $m_X \gtrsim 6$\,MeV for $\eps_A=0$. 
When we include the effects of the non-vanishing kinetic mixing, 
the lower bound becomes 10\,MeV for $\eps_A\simeq7.2\times10^{-6}$ 
corresponding to $g_X=5\times10^{-4}$. 
For further details of the contribution to $N_{\rm eff}$ and the cosmological implication of the light U(1)$_{\mutau}$ gauge boson, see {\it e.g.} Ref.~\cite{Escudero:2019gzq}. 
We conservatively use $m_X \gtrsim 6$\,MeV as the lower mass bound in this paper. 

\section{Dark matter physics}
\label{sec:DM}

In this section, we study the DM thermal production and constraints from direct detection and CMB observations. 

\subsection{Relic density}

The DM thermal relic abundance is determined by solving the Boltzmann equation, 
\begin{equation}
\frac{dn_d}{dt} + 3 H n_d = - \frac{\VEV{\sigma_{\rm ann} v}}{2} \left[ n_d^2 - (n_d^{\rm eq})^2 \right] ,
\end{equation}
where $H$ denotes the Hubble parameter, 
$n_d^{\rm (eq)} = n_\psi^{\rm (eq)} + n_{\bar{\psi}}^{\rm (eq)}$ the total (equilibrium) number density of DM, 
$\VEV{\sigma_{\rm ann} v}$ the DM total annihilation cross section averaged over the thermal distributions of the initial state. 
Study on precise calculation of the thermal relic density~\cite{Steigman:2012nb,Saikawa:2020swg} 
indicates the canonical value of the cross section to explain the observed DM abundance,
\begin{equation}
\VEV{\sigma_{\rm ann} v}/2 \simeq 3\times10^{-26}\,{\rm cm^3/s} ,
\end{equation}
for DM mass ranging from MeV to 100\,TeV. 
In this paper, we employ {\tt micromegas\_4\_3\_5}~\cite{Belanger:2014vza} to calculate the DM thermal abundance. 

There are two relevant annihilation processes, 
$\psi\bar{\psi} \to XX, f \bar{f}$ ($f=\mu,\tau,\nu_\mu,\nu_\tau$) in this model. 
The dominant annihilation mode depends on the mass spectrum of $\psi$ and $X$. 
If $\psi$ is lighter than $X$, the only possible annihilation is $\psi\bar{\psi} \to f \bar{f}$ through the $s$-channel $X$ boson exchanging. 
The cross section is given by 
\begin{equation}
(\sigma v)_{f\bar{f}} = \left\{
\begin{array}{ll} 
\displaystyle \frac{q_\psi^2 g_X^4}{2\pi} \frac{(2m_\psi^2 + m_f^2) (1-m_f^2/m_\psi^2)^{1/2}}{(4m_\psi^2-m_X^2)^2+m_X^2 \Gamma_X^2} & ~~ (f=\mu,\tau), \\
\displaystyle \frac{q_\psi^2 g_X^4}{2\pi} \frac{m_\psi^2}{(4m_\psi^2-m_X^2)^2+m_X^2 \Gamma_X^2} & ~~ (f=\nu_\mu,\nu_\tau), 
\end{array}
\right.
\end{equation}
where $\Gamma_X$ denotes the total decay width of $X$ and 
we only keep the partial $s$-wave. 
Since the gauge coupling $g_X$ is as small as ${\cal O}(10^{-4})$, 
the resulting cross section is too small to achieve the correct relic abundance for $q_\psi={\cal O}(1)$. 
The sufficiently large cross section is realized for $q_\psi\gg1$ or 
in the resonant region with $m_\psi \simeq m_X/2$. 
A comprehensive study of the resonant production has been made 
on a benchmark point $m_\psi = 0.45\,m_X$~\cite{Foldenauer:2018zrz}, 
including various experimental constraints. 
We study another option of exploiting the large DM charge as well as the resonant case. 

If DM is heavier than $X$, the other annihilation channel $\psi\bar{\psi} \to XX$ opens. 
The cross section is 
\begin{equation}
(\sigma v)_{XX} = \frac{(q_\psi g_X)^4}{4\pi m_\psi} \frac{(m_\psi^2-m_X^2)^{3/2}}{(2m_\psi^2 - m_X^2)^2} .
\end{equation}
This is proportional to $q_\psi^4$, while the cross section for $\psi\bar{\psi} \to f\bar{f}$ to $q_\psi^2$. 
Since there is no resonance enhancement in this mass regime, 
$q_\psi$ has to be large for the successful thermal production. 
Therefore, the DM abundance is produced mostly by the $\psi\bar{\psi} \to XX$ process.

\subsection{Direct detection constraints}

Elastic scattering of DM with a nucleus and an electron is severely constrained by 
direct detection experiments. 
The scattering occurs via one-loop kinetic mixing with photon $\epsilon_A$. 
Since the mixing with the $Z$ boson is suppressed by a factor of $m_X^2/m_Z^2$, 
compared with the leading contribution, we neglect it here. 

The cross section of the spin-independent DM-nucleus scattering is given by 
\begin{equation}
\sigma_N = \frac{\mu_N^2}{\pi} \frac{Z^2 \epsilon_A^2 e^2 q_\psi^2 g_X^2}{m_X^4} 
\end{equation}
where $\mu_N=m_N m_\psi/(m_\psi + m_N)$ is the reduced mass of DM and a nucleus 
with $m_N$ being the nucleus mass, and $Z$ the atomic number of the nucleus. 
This scattering is very similar to the one via the photon exchanging induced by the DM charge radius, 
$b_\psi = \eps_A q_\psi g_X / m_X^2$. 
Thus, the cross section is proportional to $Z^2$ rather than $A^2$ with $A$ being the atomic mass. 
The most stringent limit on the cross section is set by 
XENON1T ($5\,{\rm GeV} \lesssim m_\psi$) \cite{Aprile:2018dbl}, 
XENON1T with ionization signals ($3\,{\rm GeV} \lesssim m_\psi \lesssim 5\,{\rm GeV}$) \cite{Aprile:2019xxb}, 
DarkSide50 ($2\,{\rm GeV} \lesssim m_\psi \lesssim 3\,{\rm GeV}$) \cite{Agnes:2018ves}, 
XENON1T with migdal effects ($0.1\,{\rm GeV} \lesssim m_\psi \lesssim 2\,{\rm GeV}$) \cite{Aprile:2019jmx} 
and a {\tt TEA-LAB} simulated experiment inspired by the DarkSide50 result 
($0.05\,{\rm GeV} \lesssim m_\psi \lesssim 0.1\,{\rm GeV}$) \cite{GrillidiCortona:2020owp}. 

Similarly, the cross section of the DM-electron scattering is given by 
\begin{equation}
\ol{\sigma}_e = \frac{\mu_e^2}{\pi} \frac{\eps_A^2 e^2 q_\psi^2 g_X^2}{(m_X^2 + \alpha^2 m_e^2)^2}
\end{equation}
where $\mu_e$ denotes the DM-electron reduced mass. 
The constraints strongly depend on the scenario if the scattering is dependent on the momentum transfer $q$ or not. 
If the $X$ boson mass is much lighter than keV, the scattering is enhanced at a small momentum transfer. 
Then, we have to take into account a DM form factor, $F_{\rm DM}=(\alpha m_e/q)^2$. 
Since we are interested in $m_X\gtrsim6$\,MeV, however, the scattering is momentum-transfer independent, so that we can use the limit with a DM form factor, $F_{\rm DM} = 1$, in the literature. 
The most stringent limit is set by 
XENON1T ($30\,{\rm MeV} \lesssim m_\psi$) \cite{Aprile:2019xxb}, 
DarkSide50 ($20\,{\rm MeV} \lesssim m_\psi \lesssim 30\,{\rm MeV}$) \cite{Agnes:2018oej}, 
XENON10 ($10\,{\rm MeV} \lesssim m_\psi \lesssim 20\,{\rm MeV}$) \cite{Essig:2012yx}, 
and SENSEI ($m_\psi \lesssim 10\,{\rm MeV}$) \cite{Barak:2020fql}.

\subsection{Indirect bounds}

DM pair-annihilates into charged leptons and neutrinos. 
These annihilations during the cosmic dark ages and the BBN era 
have impacts on the ionization history and $N_{\rm eff}$, 
bringing us cosmological bounds. 
We will give brief comments on these bounds in the following. 

\subsubsection*{A. CMB}

Annihilation of DM into charged particles and photons increases the ionization fraction in the post-recombination era, modifying the CMB anisotropies and in turn providing the strong limit on the cross section. 
According to Ref.\cite{Slatyer:2015jla,Leane:2018kjk}, 
we adopt a conservative limit on the thermal averaged cross section into charged final states, 
$(\sigma v)_{\rm charged}/(2m_\psi) \leq 5.1 \times 10^{-27}\,{\rm cm^3\,s^{-1}\,GeV^{-1}}$\footnote{Since we consider Dirac DM with the symmetric relic, the left-hand side is divided by 2.}. 
It suggests that there is a lower mass bound $m_\psi \gtrsim 5$\,GeV 
if the annihilation into charged particles is only responsible for the thermal production. 

Let us see the impact of the CMB limit on the model. 
For $m_\psi \leq m_X$, DM can only annihilate to muon, tau and neutrinos. 
Since the cross section for these three processes is almost same, 
$(\sigma v)_{\mu\bar{\mu},\tau\bar{\tau}} \simeq 10^{-26}\,{\rm cm^3/s}$ is suggested if kinematically possible, 
excluding the DM mass up to several GeV. 
Thus, the CMB observation totally excludes the region where $m_\mu < m_\psi$. 

For $m_\psi > m_X$, the annihilation is dominated by $\psi\bar{\psi} \to XX$ followed by $X\to\nu\bar{\nu}$. 
The cross section for the annihilation into charged states is smaller 
by a factor of $(1/q_\psi^{2})$ than $\psi\bar{\psi} \to XX$. 
In this case, there is less significant energy release from DM annihilation into electron and photon. 
The CMB bound is much weaker than that for $m_\psi \leq m_X$. 
Nonetheless, the CMB observations partly limit the parameter space. 
We shall estimate it by considering $m_X \ll m_\psi$ for simplicity. 
The cross section of $\psi\bar{\psi}\to\mu\bar{\mu}$ is related to that of $\psi\bar{\psi}\to XX$ as
\begin{equation}
(\sigma v)_{\mu\bar{\mu}} = \frac{\left[1+m_\mu^2/(2m_\psi^2)\right] \left(1-m_\mu^2/m_\psi^2\right)^{1/2}}{q_\psi^2} \times (\sigma v)_{XX}. 
\label{fig;cs}
\end{equation}
The canonical cross section is $(\sigma v)_{XX}/2 \simeq 5 \times 10^{-26}\,{\rm cm^3/s}$ 
for Dirac DM~\cite{Steigman:2012nb,Saikawa:2020swg} in the sub-GeV region 
where the direct detection bound will be avoided. 
Plugging the canonical value in Eq.(\ref{fig;cs}), the CMB bound reads
\begin{equation}
q_\psi^2 \gtrsim 9.8 \left(\frac{{\rm GeV}}{m_\psi}\right) \left(1+\frac{m_\mu^2}{2m_\psi^2}\right) \left( 1-\frac{m_\mu^2}{m_\psi^2}\right)^{1/2} .
\label{eq;cmb}
\end{equation}
It follows from this equation that $q_\psi \lesssim 6.9$ is excluded for $m_\psi=200$\,MeV. 
We illustrate the CMB bound expressed by Eq.(\ref{eq;cmb}) in Fig.~\ref{fig;allowed} (right).

\subsubsection*{B. Effective neutrino number $N_{\rm eff}$}

The annihilation into neutrinos reheats the SM plasma when DM becomes non-relativistic. 
If it occurs after the neutrino decoupling, 
only the neutrino is reheated and the neutrino-to-photon temperature ratio increases, 
resulting in the higher expansion rate at the BBN than the standard scenario. 
The expansion rate at the BBN, together with the baryon-to-photon ratio, affects the primordial abundances of helium and deuteron. 

\begin{figure}[t]
\includegraphics[width=0.5\textwidth]{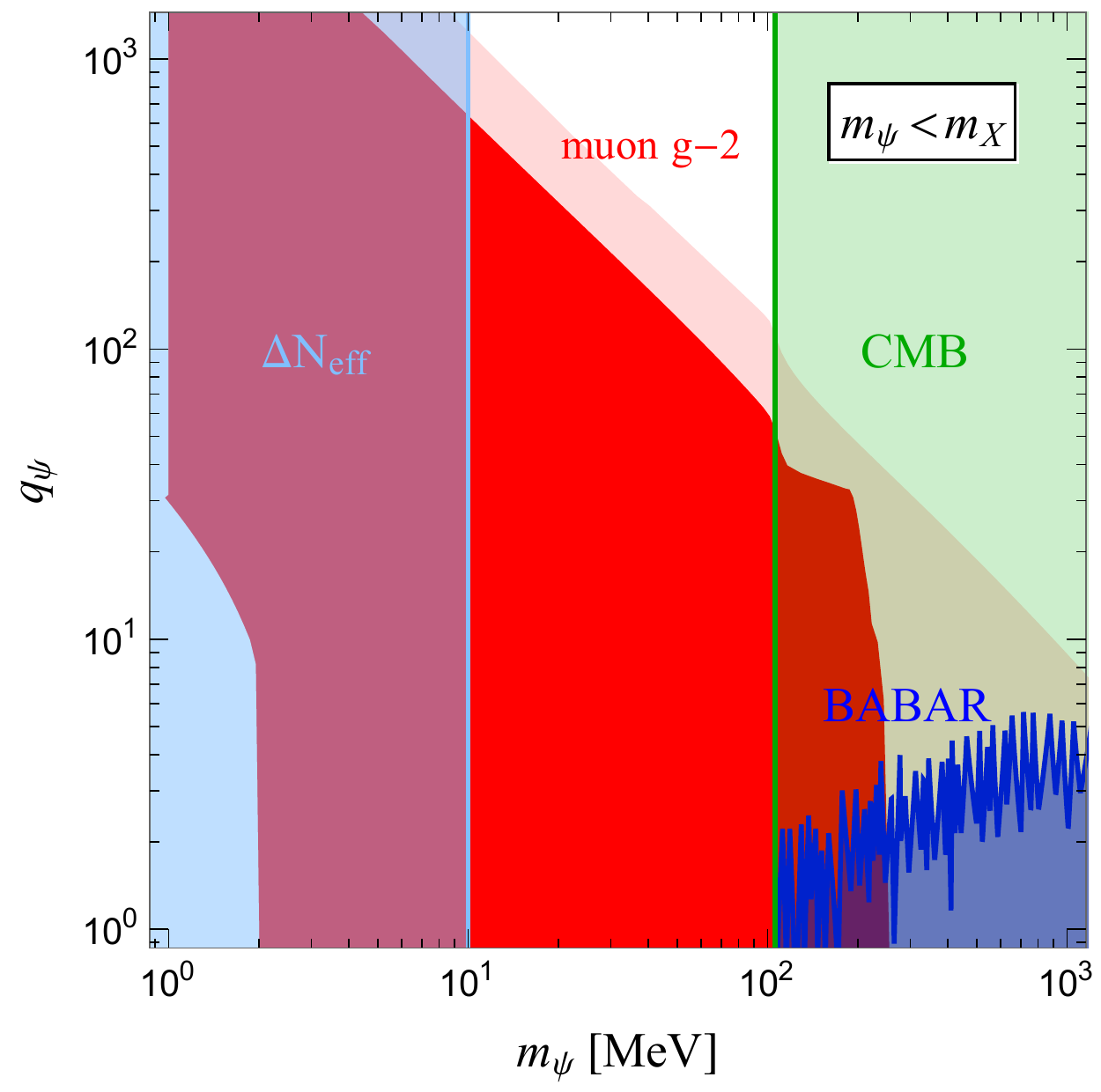}
\includegraphics[width=0.5\textwidth]{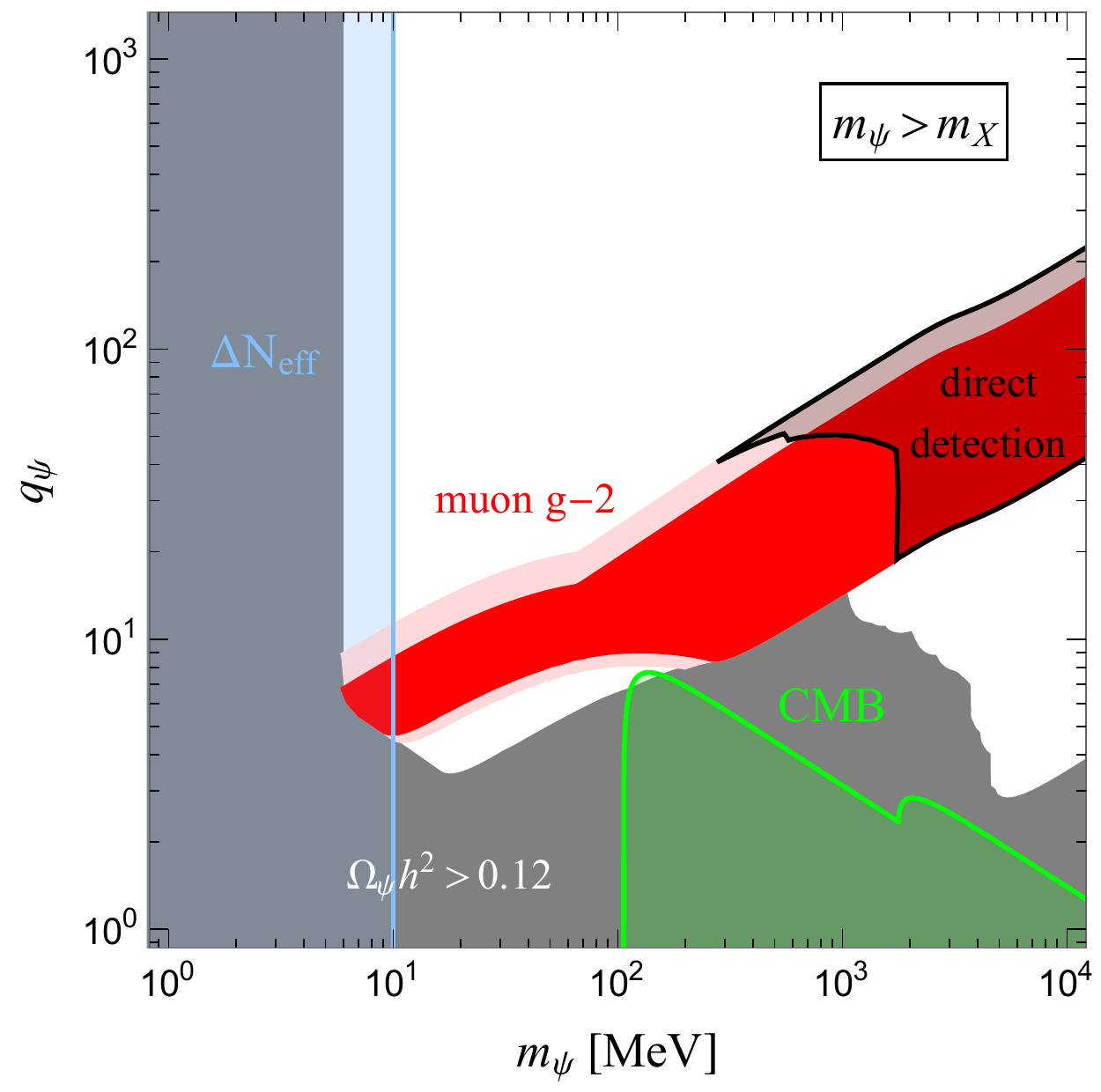}
\caption{
The allowed parameter region in the plane of the DM parameters.
Dark (light) red region is favored the muon $g-2$ at $1\sigma \, (2\sigma)$ level.
Shaded regions are excluded by the CMB (green), $\Delta N_{\rm eff}$ (light blue) and direct detection (black) and BABAR (blue) experiments. 
Dark gray region is disfavored by the DM overabundance ($\Omega_\psi h^2 > 0.12$).  
}
\label{fig;allowed}
\end{figure}

The change of the expansion rate is rendered in the effective neutrino number $N_{\rm eff}$. 
A significant deviation of $N_{\rm eff}$ from the standard value $N_{\rm eff,SM} = 3.046$ ~\cite{Mangano:2005cc}
is faced with the precise Planck observations. 
In \cite{Boehm:2013jpa,Nollett:2014lwa,Heo:2015kra,Sabti:2019mhn}, 
the authors calculate the increase of $N_{\rm eff}$ by a relic particle coupled to neutrinos, 
assuming that the particle is only in the equilibrium with the neutrinos during the BBN. 
They obtain the lower mass bound of Dirac DM $m_\psi \gtrsim 10$\,MeV, 
which is drawn in Fig.~\ref{fig;allowed} with light blue.

\subsection{Allowed region}

Figure \ref{fig;allowed} shows the allowed region in the ($m_\psi, q_\psi$) plane. 
The left panel corresponds to $m_\psi < m_X$, while the right panel to $m_\psi > m_X$. 
In both panels, we scan two parameters, $m_X$ and $g_X$, 
such that the experimental and cosmological limits given in Sec.~\ref{sec:model} are avoided. 
The DM observed abundance, $\Omega_{\rm CDM} h^2 = 0.12$, can be explained 
in all the regions in the plots, except for the dark gray region where $\Omega_\psi h^2 > 0.12$ is predicted. 
We also highlight the $1\sigma$ ($2\sigma$) muon $g-2$ favored region with the (light) red band. 

In Fig.~\ref{fig;allowed} (left) with $m_\psi < m_X$, 
the muon $g-2$ explanation and DM production are simultaneously achieved 
in a wide DM mass range. 
In particular, we find that the non-resonant DM production is realized for $q_\psi \gtrsim {\cal O}(10)$, 
otherwise the large resonant enhancement is needed. 
We also show the excluded regions from the CMB (green) and $N_{\rm eff}$ (light blue). 
These restrict the DM mass to be $10\,{\rm MeV} \lesssim m_\psi \lesssim 100\,{\rm MeV}$. 
Note that the BABAR limit (blue) depends on $q_\psi$ for $m_\psi < m_X/2$. 
We estimate this limit by rescaling the announced BABAR limit~\cite{TheBABAR:2016rlg} 
with the branching ratio: 
$g_X \to g_X \cdot \{ {\rm Br}_{X\to\mu\bar{\mu}}|_{q_\psi} / {\rm Br}_{X\to\mu\bar{\mu}}|_{q_\psi=0} \}^{1/2}$. 
Naively speaking, the limit is weakened by a factor of $q_\psi$, compared to the announced one. 
The BABAR limit excludes the region of $q_\psi \lesssim 5$, 
although it is overlapped with the CMB limit. 

It is obvious from Fig.~\ref{fig;allowed} (right) that 
allowing the large $q_\psi$ opens a new mass regime, $m_\psi > m_X$, 
that has not been pointed out in the previous study. 
The heavy DM mass, even ${\cal O}(100\,{\rm GeV})$, is allowed in this regime. 
The leading constraints in the heavy mass region are set by the direct detection experiments. 
Given the current results, the DM mass should be lighter than 2\,GeV to resolve the muon $g-2$ anomaly. 
In contrast, if we do not require the muon $g-2$ explanation, the direct detection limit can be very weak, 
since we can consider the heavy $X$ boson outside the light red band in Fig.~\ref{fig;Xboson}. 
The gain of the $X$ boson mass does not affect the DM production much, 
while it considerably reduces the elastic scattering with nuclei and electrons, thereby weakening the direct detection bound\footnote{
The leading constraint on the heavy $X$ boson is {\it e.g.} from a CMS search~\cite{Sirunyan:2018nnz}, 
but it is weaker than the constraints on the light $X$ boson. 
Thus, we have no difficulty in making the $X$ boson heavy enough to avoid the direct detection limit.}.
As far as the indirect bounds are concerned, 
the $N_{\rm eff}$ bound disfavors $m_\psi \lesssim 10$\,MeV, 
while the CMB gives a complementary constraint in the ${\cal O}$(100\,MeV) region, 
but only a small part is covered yet. 
Altogether, it is $10\,{\rm MeV} \lesssim m_\psi \lesssim 2$\,GeV where 
the muon $g-2$ anomaly and DM are both addressed. 
One may wonder if the large DM charge will generate a sizable Sommerfeld enhancement at the freeze-out or in indirect searches. 
We have confirmed, however, that there is no sizable Sommerfeld effect in the parameters space we consider.

It has turned out so far that 
allowing the large DM charge opens two new DM production regimes; 
(i) the non-resonant production for $m_\psi < m_X$ and 
(ii) the secluded production for $m_\psi > m_X$. 
These are overlooked possibilities in the previous study. 
Given the new phenomenological possibilities, 
it is natural to explore the signals of DM in this new parameter space. 
In the next section, we will study the indirect signals of this DM candidate 
at neutrino telescope experiments.

\section{Signature with Neutrino Telescope}
\label{sec:indirect}

We have seen that the large DM charge revives the successful thermal production 
without the help of the resonance enhancement. 
As a by-product of the revival, the model predicts an excess of the neutrino flux from the galactic space. 
In this section, we discuss the sensitivity of the SK to the neutrino flux excess and the future prospect at the HK. 

There are some dedicated studies of neutrino signatures of annihilating DM 
in the framework of simplified models
\cite{Yuksel:2007ac,PalomaresRuiz:2007eu,Primulando:2017kxf,Campo:2017nwh,Campo:2018dfh,Klop:2018ltd,Arguelles:2019ouk,Bell:2020rkw}. 
In these works, it is assumed that DM directly annihilates into a pair of neutrinos. 
The resulting monochromatic neutrino fluxes are compared with the results of 
searches for unknown extraterrestrial neutrino sources~\cite{Bellini:2010gn,Collaboration:2011jza}, 
SRN searches~\cite{Malek:2002ns,Bays:2011si,Zhang:2013tua} and 
atmospheric neutrino measurements~\cite{Richard:2015aua}, 
to derive upper limits on the annihilation cross sections. 
In the following, 
we perform an independent analysis for the DM signals measured at the SK, 
assuming that the extra neutrino flux originates from 
the direct annihilation $\psi\bar{\psi} \to \nu\bar{\nu}$ or 
the secluded annihilation $\psi\bar{\psi} \to XX \to 2\nu 2\bar{\nu}$ in the Milky Way. 
We also note that the analysis in this section is applicable 
as far as the event topologies are same only by changing the model dependent flux normalization. 

\subsection{Neutrino flux from DM annihilation}

We here assume that neutrinos are Dirac particles for definiteness, 
but the result will be unchanged even if these are Majorana particles. 
The expected electron (anti-)neutrino flux at the detector from DM pair annihilation in the galactic halo is expressed by 
\beq
\frac{d\Phi_{\nu_e (\bar{\nu}_e)}}{dE_\nu} = \frac{1}{4\pi} \sum_i \frac{\VEV{\sigma v}_i}{4m_\psi^2} 
\,\kappa\, 
\frac{dN_i}{dE_\nu} J_{\Delta\Omega},
\label{eq;nuflux}
\eeq
where $\VEV{\sigma v}_i$ denotes the annihilation cross section into a final state $i$, 
$\kappa$ is a model dependent constant which characterizes electron-neutrino flavor fraction, 
$dN_i/dE_\nu$ the neutrino spectral function for the final state $i$, 
$J_{\Delta\Omega}$ the astrophysical $J$-factor which is given by the line-of-sight (l.o.s) integral of the DM density square. 
In the ($b,l$) coordinate, the $J$-factor is expressed by 
\beq
J_{\Delta\Omega} = \int_{b_{\rm min}}^{b_{\rm max}} db\,\cos b \int_{l_{\rm min}}^{l_{\rm max}} dl 
\int_0^{s_{\rm max}} ds \, \rho(r(s,b,l))^2 , 
\eeq
where $\rho(r)$ is the DM density profile in the galactic halo. 
The galactocentric distance $r$ is expressed in terms of the l.o.s distance $s$, 
\begin{equation}
r(s,b,l) = \sqrt{s^2+r_\odot^2 - 2 s\, r_\odot  \cos b \cos l} ,
\end{equation}
with $r_\odot = 8.5$\,kpc being the distance between the galactic center (GC) and the solar system. 
We define the direction of the GC as $b=0^\circ$ and $l=0^\circ$. 
The size of the l.o.s distance is restricted by the DM halo size, whose maximum is given by
\begin{equation}
s_{\rm max}(b,l) = \sqrt{R_{\rm halo}^2 - r_\odot^2 + r_\odot^2 \cos^2b\,\cos^2l} + r_\odot \cos{b}\,\cos{l}.
\end{equation}
We take the DM halo size $R_{\rm halo}=40$\,kpc. 
Note that the $J$-factor is insensitive to the halo size parameter 
as far as $R_{\rm halo} \geq 30$\,kpc. 
Since the DM profile is spherical symmetric, 
we define the range of $(b,l)$ as $0^\circ \leq b \leq 90^\circ$ and $0^\circ \leq l \leq 180^\circ$ 
and then multiply the result by four. 
We consider the Navarro-Frenk-White (NFW) profile~\cite{Navarro:1996gj,Vertongen:2011mu}, 
\begin{equation}
\rho(r) = \rho_s \left(\frac{r_s}{r}\right) \frac{1}{1+(r/r_s)^2} ,
\end{equation}
where $r_s=20$\,kpc is the scale radius and 
the density profile is normalized with $\rho(r=r_\odot)=0.4\,{\rm GeV/cm^3}$. 
The resulting all-sky $J$-factor is $J_{\Delta \Omega} \simeq 1.5 \times 10^{23}\,{\rm GeV^2/cm^5}$. 
As an example for calculating the model dependent constant $\kappa$, 
if the neutrino flavor ratio at source is $\nu_e : \nu_\mu : \nu_\tau = 0 : 1 : 1$, 
then the flavor ratio at the detector is $1 : 2 : 2$. 
Thus, $\kappa=1/5$ is obtained for the U(1)$_{\mutau}$ DM model. 

The neutrino flux exhibits a distinctive feature. 
There are two important annihilation modes, which are $\psi\bar{\psi} \to \nu\bar{\nu}, XX$. 
The former is dominant for $m_\psi \leq m_X$, while the latter for $m_\psi > m_X$. 
In the former annihilation, the neutrino spectrum is monochromatic at $E_\nu=m_\psi$ 
in the center-of-mass frame of the annihilation. 
Since DM is almost at rest in the galactic halo, 
the monochromatic feature is maintained in the lab frame that is the rest frame of the DM halo. 
When the annihilation occurs in the Milky Way, the redshift is negligible. 
Therefore, the neutrino spectrum at the detector keeps the monochromatic form, 
$dN/dE_\nu \propto \delta(E_\nu-m_\psi)$. 
This spiky spectrum is easy to be disentangled from the smooth background spectrum, 
providing a good sensitivity. 

In the secluded annihilation $\psi\bar\psi \to XX \to 2\nu 2\bar\nu$, the neutrino spectrum takes another characteristic form. 
The spectral feature from such a cascade annihilation has been studied in \cite{Garcia-Cely:2016pse} for an intermediate state with an arbitrary spin, and the authors have shown that the spectral shape in general depends on the polarization of the intermediate state. 
In what follows, we restrict ourselves to an annihilation process $\psi\bar\psi \to V V \to 2\nu 2\bar\nu$ with a vector boson $V$ (not necessarily be the $X$ boson), and briefly review the spectral form. 
In the rest frame of the decaying $V$, 
each of emitted neutrinos has a monochromatic spectrum at $E_\nu^\prime = m_V$. 
In the lab frame, the neutrino energy is boosted and reads 
\begin{equation}
E_\nu = \frac{m_\psi}{2} \frac{r_V^2}{1 - \cos\theta \sqrt{1 - r_V^2} },
\end{equation}
where $r_V=m_V/m_\psi$ and $\theta$ is the angle between momenta of the parent $V$ boson and the emitted neutrino in the lab frame. 
There is a sharp kinematical cut of the neutrino energy. 
The maximum (minimum) neutrino energy corresponds to $\theta = 0^\circ \, (180^\circ)$. 
Since the emission angle $\theta$ determines the neutrino energy in the lab frame, 
we can obtain the neutrino energy spectrum once we know the angular distribution of the neutrino emitted by the $V$ boson decay. 
If the $V$ boson produced in $\psi\bar\psi\to VV$ is unpolarized, the neutrino emission is isotropic and has no angular distribution. 
In this case, the neutrino spectrum has no energy dependence between the kinematical endpoints, {\it i.e.} it takes the so-called box shape~\cite{Ibarra:2012dw,Ibarra:2013eda}, 
\begin{equation}
\frac{dN}{dE_\nu} = \frac{2}{m_\psi (1-r_V^2)^{1/2}} \Theta(y - y_-) \Theta(y_+ - y) ,
\label{eq;box}
\end{equation}
where $y=E_\nu/m_\psi$ and $y_\pm = \left( 1 \pm \sqrt{1-r_V^2} \right)/2$ and $\Theta(y)$ is the Heaviside theta function. 
On the other hand, if the produced $V$ boson is polarized, the neutrino emission has an angular distribution and hence the resulting neutrino spectrum has an energy dependence. 

In general, the neutrino spectrum in the $\psi\bar\psi\to VV \to 2\nu 2\bar\nu$ process is expressed by 
\begin{equation}
\frac{dN}{dE_\nu} = \frac{1}{m_\psi} \sum_{m,n} {\rm Br}_{m,n} \left[ f_m (y) + f_n(y) \right] ,
\label{eq;nuspectrum}
\end{equation}
where the summation runs over $V$ boson helicity $m,n =+1,\,0,\,-1$ and 
${\rm Br}_{m,n}$ denotes the production branching fraction for each helicity state, 
\begin{equation}
{\rm Br}_{m,n} = \frac{\sigma v (\psi\bar\psi \to V_m V_n)}{\sum_{m,n} \sigma v (\psi\bar\psi \to V_m V_n)} .
\end{equation}
The spectral functions $f_m(y)$ are in the form of 
\begin{align}
f_0 (y) & = \frac{3}{2} \frac{4y-4y^2-r_X^2}{(1-r_X^2)^{3/2}} \Theta(y-y_-) \Theta(y_+-y), \\
f_{\pm1} (y) & = \frac{3}{4} \frac{\Theta(y-y_-) \Theta(y_+-y)}{(1-r_X^2)^{3/2}} \nonumber\\
& \quad \times \left[ 2-4y+4y^2-r_X^2 \pm 2 (C_{+1}-C_{-1}) (2y-1) \sqrt{1-r_X^2} \right] ,
\end{align}
where $C_m$ are model-dependent coefficients that 
characterize how the intermediate state with the different polarizations couples to the decay products. 
For example, when the vector boson couples to the final state fermions as 
$\ol{f} \left( g_R \gamma^\mu P_R + g_L \gamma^\mu P_L \right) f^\prime V_\mu$, we find 
\begin{align}
C_0 & \simeq \frac{m_f^2 + m_{f^\prime}^2}{2m_V^2} + \frac{2g_Lg_R}{g_L^2+g_R^2} \frac{m_f m_{f^\prime}}{m_V^2} , \\
C_{+1} & \simeq \frac{g_R^2(1-C_0)}{g_L^2+g_R^2} + \frac{2g_Lg_R(g_L^2-g_R^2)}{(g_L^2+g_R^2)^2} \frac{m_f m_{f^\prime}}{m_V^2} ,\\
C_{-1} & \simeq \frac{g_L^2(1-C_0)}{g_L^2+g_R^2} - \frac{2g_Lg_R(g_L^2-g_R^2)}{(g_L^2+g_R^2)^2} \frac{m_f m_{f^\prime}}{m_V^2} , 
\end{align}
in the leading order in $m_f/m_V$ or $m_{f^\prime}/m_V$~\cite{Garcia-Cely:2016pse}. 
Here, we would like to mention the relation, 
\begin{equation}
\frac{1}{3} \sum_m f_m (y) = \frac{\Theta(y-y_-) \Theta(y_+-y)}{(1-r_V^2)^{1/2}} .
\end{equation}
It follows from this relation that if the branching fraction is helicity-independent, {\it i.e.} the $V$ boson is unpolarized, then we exactly obtain the box-shape spectrum Eq.(\ref{eq;box}). 

\begin{figure}[t]
\centering
\includegraphics[width=0.6\textwidth]{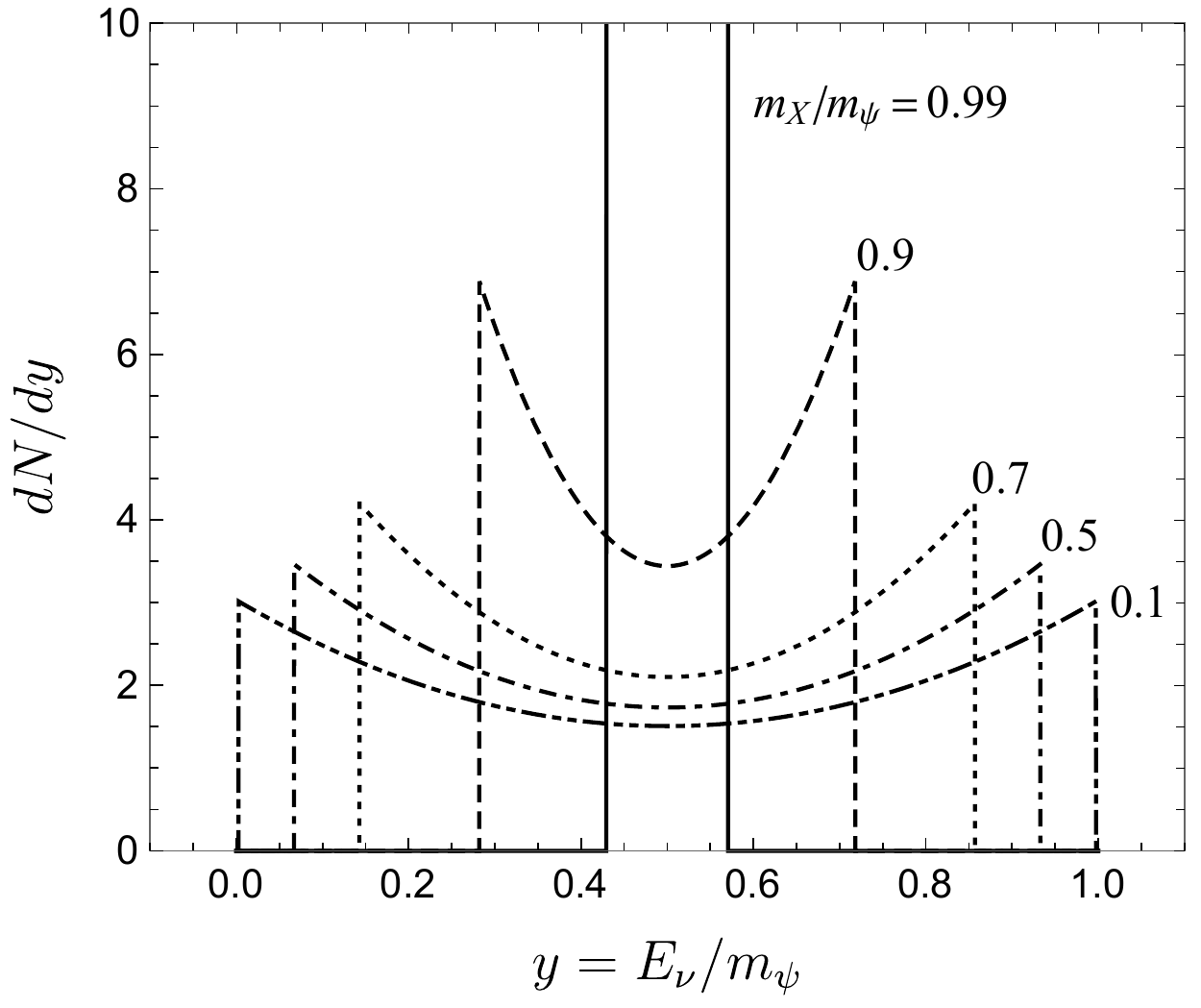}
\caption{The shape of the neutrino spectrum for the $\psi \bar{\psi} \to XX \to2\nu2\bar{\nu}$ annihilation.}
\label{fig;spectrum}
\end{figure}

Now let us return to our model. 
The $X$ boson mostly couples to the left-handed neutrinos which can be regarded as massless, so that $C_0 \simeq C_{+1} \simeq 0$ and $C_{-1} \simeq 1$. 
Further, the polarization of the $X$ boson produced in the annihilation $\psi\bar\psi \to X_m X_n$ is purely transverse, 
\begin{equation}
{\rm Br} = {\rm diag} (1/2,0,1/2) .
\end{equation}
As a result, the neutrino spectrum is in the form of a bowl-shape, 
\begin{equation}
\frac{dN}{dE_\nu} = \frac{2}{m_\psi} \frac{3(2-4y+4y^2-r_X^2)}{4(1-r_X^2)^{3/2}} \Theta(y-y_-) \Theta(y_+-y) .
\label{eq;transverse}
\end{equation}
In Fig.~\ref{fig;spectrum}, 
we show the neutrino spectrum Eq.(\ref{eq;transverse}), which is symmetric with respect to $E_\nu = m_\psi/2$. 
The width of the spectrum is determined solely by kinematics and is given by $\sqrt{1 - r_V^2}$. 
As $\psi$ and $X$ are more degenerate, the width is narrower and the height is taller. 

We would like to remark on the DM model dependence of the neutrino spectrum from the secluded annihilation. 
For example, if DM is a complex scalar charged under the U(1)$_{\mutau}$ gauge group as considered in \cite{Garani:2019fpa}, 
the annihilation into the longitudinally polarized $X$ boson is nonvanishing and 
the branching fraction depends on the mass degeneracy of DM and $X$, 
\begin{equation}
{\rm Br} \propto {\rm diag} \left( 1,\, \frac{r_X^4}{(2-r_X^2)^2} ,\, 1 \right) .
\end{equation}
With a small $r_X$, the annihilation into the longitudinal polarization is sub-leading and the spectral form is similar to the bowl-shape, 
while all polarizations contribute equally with $r_X \simeq 1$ and the spectrum gets close to the box-shape. 
Moreover, the annihilation into the longitudinal polarization can be even dominant 
if DM is sizably coupled to the Nambu-Goldstone mode of the U(1)$_{\mutau}$ Higgs field. 
In this case, the spectrum is like an upside-down bowl. 
With the current and even future-planned precision, however, 
it will be difficult to distinguish the difference and these spectra will look practically the same shape in experiments. 
It might be interesting if any effort in future would realize good enough capability to discern the difference of the spectra.

\subsection{Analysis}

We explain our analytical method. 
We try to reinterpret the result of the 2,853 days SK search for SRN~\cite{Bays:2011si}, 
to derive the limits on the DM annihilation cross section. 
The signal in this experiment is an electron and a positron produced in the reaction of  
the inverse beta decay (IBD) ($\bar{\nu}_e + p \to e^+ + n$) and 
the neutrino absorption by Oxygen in the charged current (CC) interactions ($\nu_e (\bar{\nu}_e) + {^{16}{\rm O}} \to e^- (e^+) + X_A$). 
The energy of the produced electron/positron is related to the initial neutrino energy, 
$E_e \simeq E_\nu -1.3$\,MeV ($\bar{\nu}_e p$), 
$E_e \simeq E_\nu -15.4$\,MeV ($\nu_e$O) and 
$E_e \simeq E_\nu -11.4$\,MeV ($\bar{\nu}_e$O). 
If no excess of the recoil events is found, one can derive the limit on the neutrino flux, 
which is in turn translated into the bound on the annihilation cross section. 

The signal region is set to $E_e=16$--88\,MeV which is divided into 18 bins with a 4\,MeV width. 
The number of the signal events measured in the $i$-th bin is expressed by 
\begin{align}
N_{i, {\rm sig}} & = N_{\rm SK}\, T_{\rm SK} \int dE_\nu \frac{d\Phi_{\nu_e}}{dE_\nu} 
\int_{E_i}^{E_{i+1}} dE_{\rm vis} 
\int dE_e \, R(E_e,E_{\rm vis}) \, \eps(E_{\rm vis}) \nonumber\\
& \quad \times \left\{ \frac{d\sigma_{\bar{\nu}_ep}}{dE_e} (E_\nu, E_e) 
+ \frac{1}{2} \left( \frac{d\sigma_{\nu_e \Ox}}{dE_e} (E_\nu, E_e) + \frac{d\sigma_{\bar{\nu}_e \Ox}}{dE_e} (E_\nu, E_e) \right) \right\},
\end{align}
where $N_{\rm SK}=1.5\times10^{33}$ denotes the number of free protons in the SK detector, 
$T_{\rm SK}$ the exposure time of SK. 
We combine SK-I (1,497\,days), SK-II (794\,days) and SK-III (562\,days) data. 
To take into account the detector efficiency and the finite energy resolution, 
we introduce the efficiency function, $\eps(E_{\rm vis})$ (figure 10 of \cite{Bays:2011si}), 
and the Gaussian-like resolution function, 
\begin{equation}
R(E_e, E_{\rm vis}) = \frac{1}{\sqrt{2\pi} \sigma} {\rm exp}\left\{ -\frac{(E_e-E_{\rm vis})^2}{2\sigma^2} \right\} ,
\end{equation}
with the width $\sigma (E_e)= 0.4\,{\rm MeV} \sqrt{E_e/{\rm MeV}} + 0.03 E_e$~\cite{PalomaresRuiz:2007eu}. 
Here, $E_e$ is the actual electron/positron energy, 
$E_{\rm vis}$ the measured energy at detector 
and $d\Phi_{\nu_e}/dE_\nu$ the neutrino flux originated from the galactic DM annihilation in Eq.(\ref{eq;nuflux}). 
We use the NLO analytical expression for the IBD cross section, $d\sigma_{\overline{\nu_e}p}/dE_e$, 
in \cite{Strumia:2003zx} 
and extract $d\sigma_{\nu_e\Ox}/dE_e$ and $d\sigma_{\overline{\nu_e}\Ox}/dE_e$ from \cite{Kolbe:2002gk,Skadhauge:2006su}. 
For the latter, we assume the electron energy is mono-energetic, 
e.g. $d\sigma_{\nu_e \Ox}/dE_\nu (E_\nu,E_e) = \sigma_{\nu_e \Ox}(E_\nu) \, \delta(E_e - E_\nu - 15.4\,{\rm MeV})$, and determine $\sigma_{\nu_e \Ox}(E_\nu)$ from Fig.2 of \cite{Skadhauge:2006su}. 

\begin{figure}[t]
\includegraphics[width=0.5\textwidth]{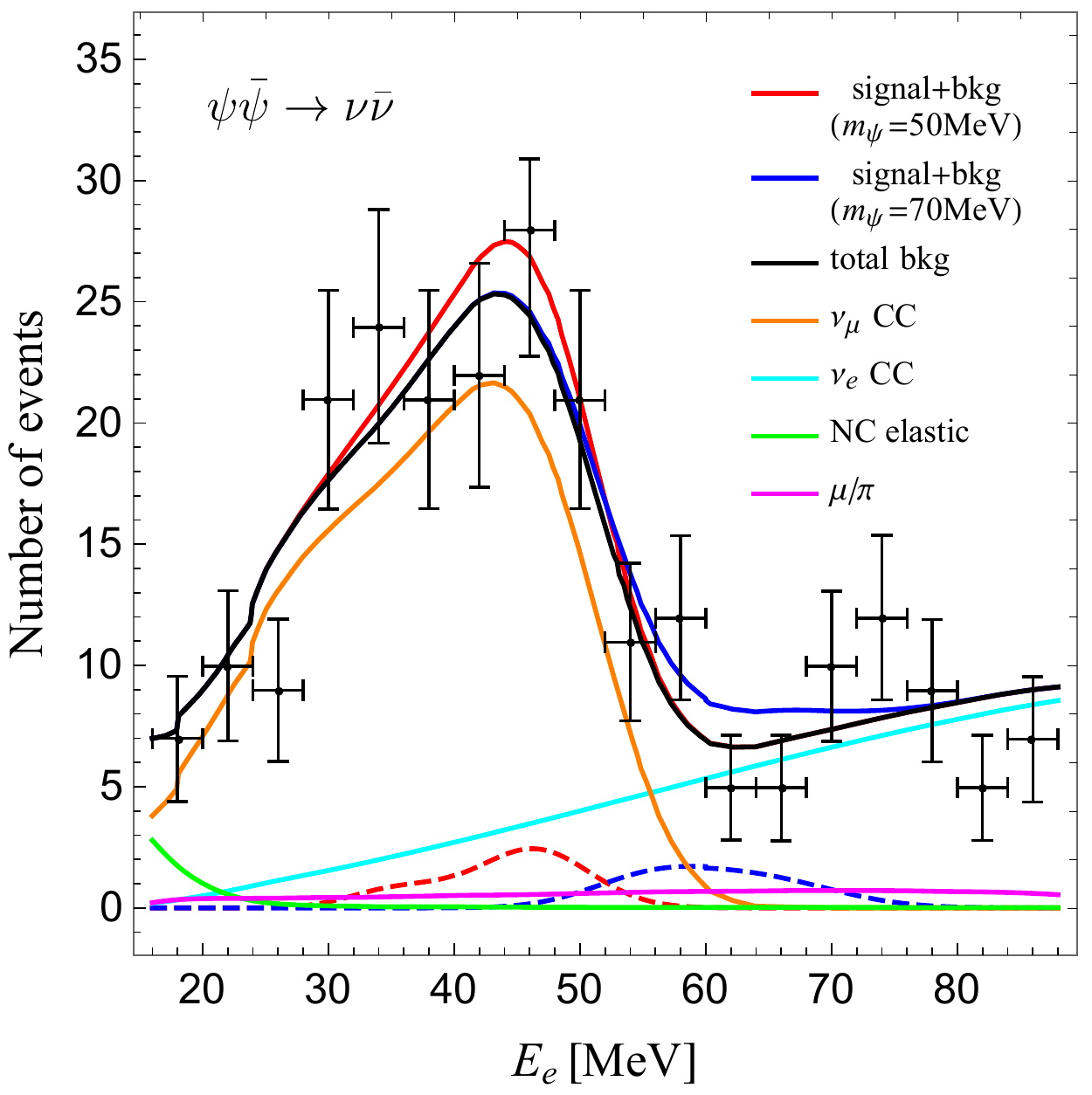}
\includegraphics[width=0.5\textwidth]{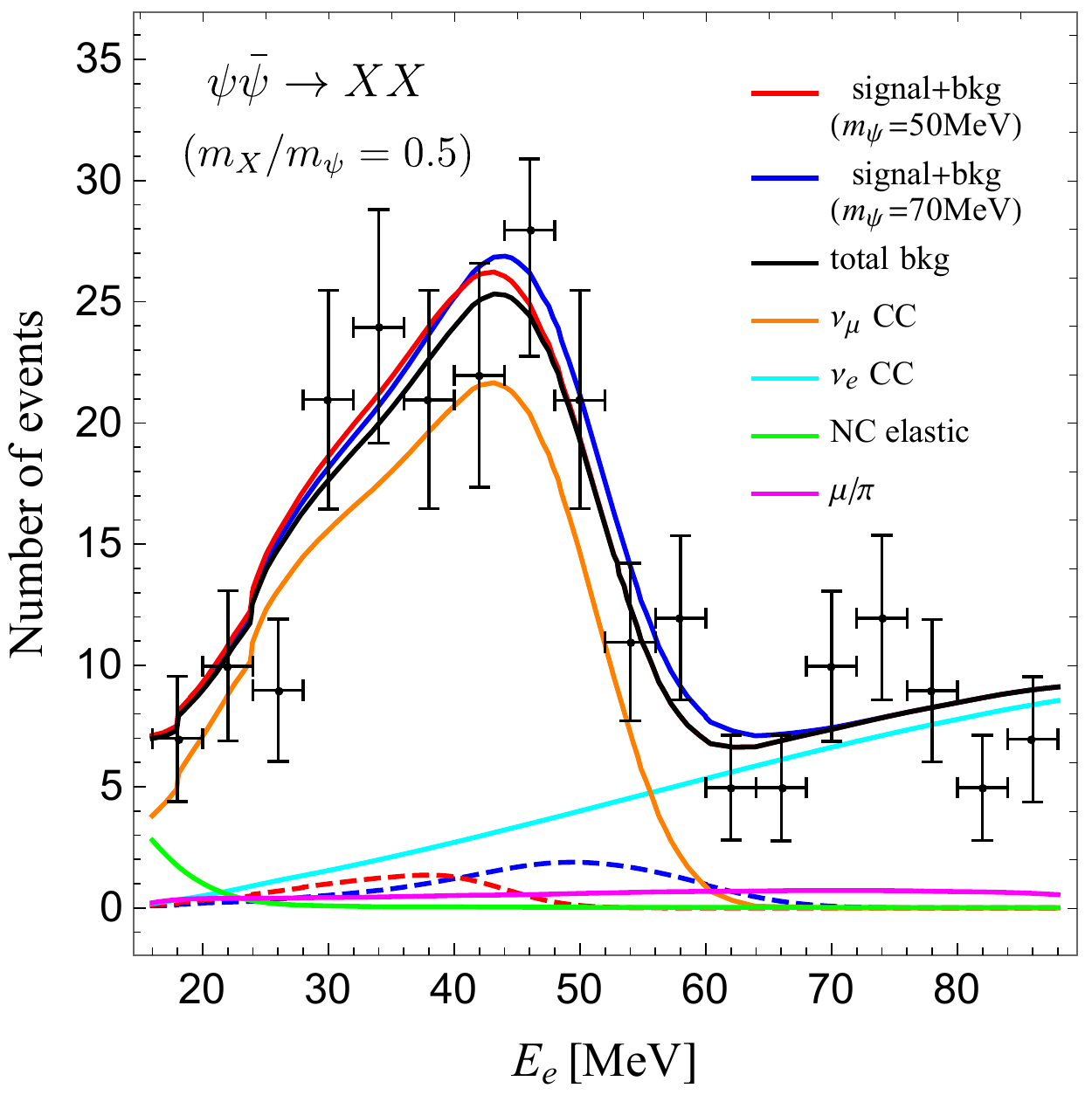}
\caption{
The DM signal events measured in the SK-I (1,497\,days) 
for $\psi\bar{\psi}\to\nu\bar{\nu}$ (left) and for $\psi\bar{\psi}\to XX$ (right). 
The cross section is fixed such that ${\rm TS}=2.71$.
Solid (dashed) red and blue curves correspond to the expected signal events with (without) backgrounds.
Black points correspond to the observed number of events in the bins with the error bars.
The other colored lines correspond to various kinds of background events that are taken from figure 14 of \cite{Bays:2011si}.
}
\label{fig;event}
\end{figure}

There are four main backgrounds in the signal region $E_e=[16,88]$\,MeV. 
The first one is an electron from the Michel decay of a muon that is produced by 
the CC interaction of the atmospheric muon neutrino in the water of the detector. 
If the momentum of the produced muon is lower than the Cherenkov threshold, 
this muon is invisible and its decay electron cannot be removed. 
This is called the invisible muon background. 
This has a kinematical edge at $E_e \simeq 60$\,MeV. 
The second one comes from the CC interaction of the atmospheric electron neutrino, 
producing the monotonically increasing events with the increasing $E_e$. 
This dominates the background above the kinematical endpoint of the invisible muon. 
The third one is an electron produced via the neutral current (NC) interaction of 
the atmospheric neutrinos. This is increased at the low energy bins. 
The last grouping of the background is due to heavy charged particles, 
pions and muons, created in the NC reactions. 
Some of them survive the pion and Cherenkov angle cuts, and enter in the signal region. 
These backgrounds are shown in Fig.~\ref{fig;event}. 

To derive the 90\% confidence level (C.L.) limit, we first introduce a likelihood function, 
\begin{equation}
{\cal L}_i (N_{i, {\rm sig}}) = \frac{(N_{i,{\rm bkg}}+N_{i, {\rm sig}})^{N_{i, {\rm obs}}}}{N_{i,{\rm obs}}\,\!!} e^{-(N_{i,{\rm bkg}}+N_{i,{\rm sig}})} ,
\end{equation}
for each bin and each SK phase. 
With this likelihood, we define test statistic (TS) as  
\begin{equation}
{\rm TS} = -2 \sum_i \ln\left(\frac{{\cal L}_i (N_{i,{\rm sig}})}{{\cal L}_i (0)}\right) ,
\end{equation}
where the summation runs over 18 energy bins and from SK-I to SK-III. 
The 90\% C.L. limit is obtained by solving ${\rm TS} \leq 2.71$. 
Since TS is a function of the DM mass and annihilation cross section ${\rm TS}(m_\psi,\sigma v)$, 
we obtain an exclusion curve in the ($m_\psi, \sigma v$) plane. 
To estimate the HK sensitivity, we consider the $374$\,kton fiducial volume and $10$\,yrs livetime 
with the same efficiency and energy resolution as the SK. 
The HK detector has an option of doping Gd to reduce the backgrounds. 
In our analysis, it is assumed that the doped Gd reduces the invisible muon background by 50\% or 80\%. 
The sensitivity curves are obtained by solving $\TS=2.71$ with $N_{\rm obs}=N_{\rm bkg}$.

\subsection{SK bounds and HK sensitivity}

We compare the DM signal events with the observed events at the SK-I (1,497\,days) in Fig.~\ref{fig;event}. 
The left panel corresponds to $\psi\bar{\psi}\to\nu\bar{\nu}$ and the right panel to $\psi\bar{\psi}\to XX$ with $m_X/m_\psi=0.5$. 
We assume 100\,\% branching fraction for the corresponding annihilation and 
fix the cross section such that ${\rm TS}=2.71$. 
The red and blue curves represent the number of events in presence of the DM annihilation. 
The dashed one is the signal and the solid one is the sum of the signal and backgrounds.

The monochromatic flux predicts the distinct event shape, 
which is peaked at $E_e \simeq m_\psi$ for 50\,MeV DM. 
The shape is broader for 70\,MeV DM, on the other hand. 
This is because there are three reactions that produce the electron/positron with different energy: 
the IBD with $E_e \simeq E_\nu - 1.3$\,MeV and $\nu_e (\bar{\nu}_e) + {^{16}}\Ox$ scattering with $E_e \simeq E_\nu - 15.4$\,(11.4)\,MeV. 
The cross sections for the latter two scatterings are still smaller at 50\,MeV than the former, 
while these become comparable at 70\,MeV. 
Thus, another peak appears at an energy 10\,MeV below the DM mass. 
The resulting event shape takes the flatter form. 
The bowl-shape flux induced by the secluded annihilation produces the broader event shape. 
The event excess in each bin is not sizable, but it equally contributes to multiple bins, 
not losing the sensitivity. 
The center of the neutrino spectrum is at a half of DM mass. 
While, the peak appears at slightly higher energy 
because the cross section grows as the energy. 

\begin{figure}[t]
\includegraphics[width=0.5\textwidth]{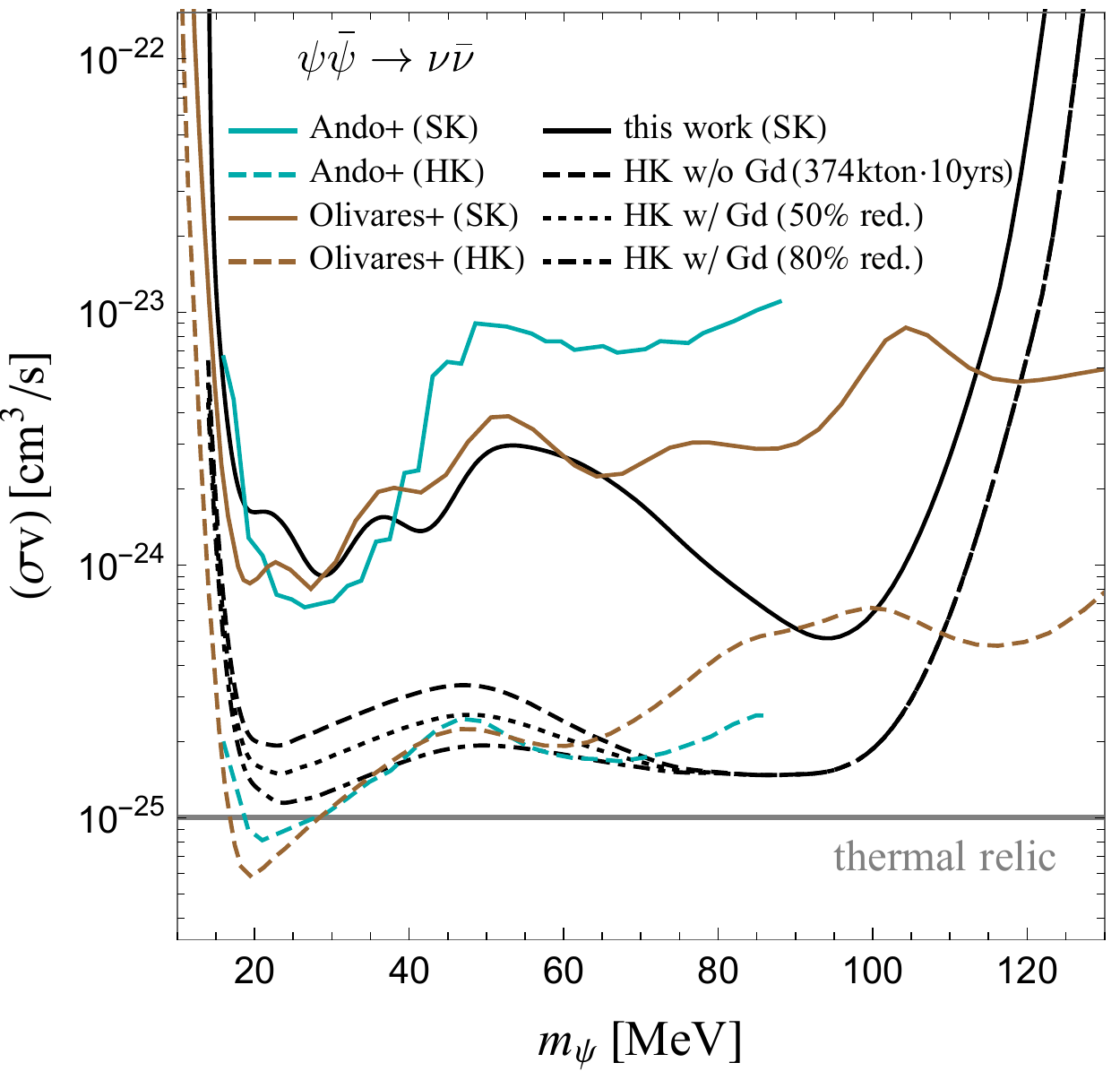}
\includegraphics[width=0.5\textwidth]{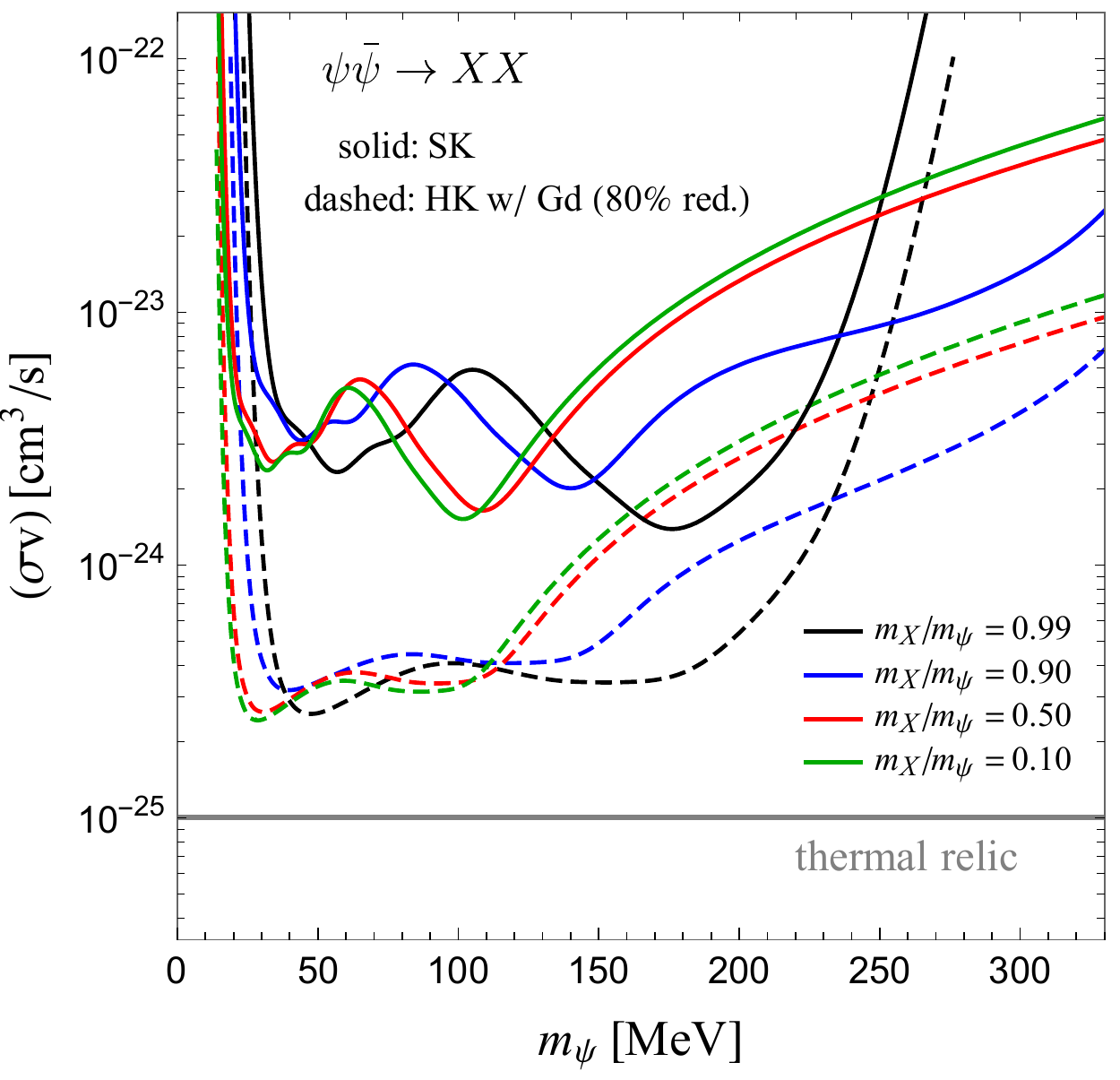}
\caption{The 90\% C.L. upper limits on and the future HK sensitivity to the annihilation cross section for $\psi\bar{\psi}\to\nu\bar{\nu}$ (left) and $\psi\bar{\psi} \to XX$ (right) cases. 
Black solid (dashed) line corresponds to the limit on the annihilation cross section from SK (HK).
The results given in \cite{Campo:2018dfh,Klop:2018ltd} are shown for reference, and are shown in turquoise and brown lines.}
\label{fig;indirect}
\end{figure}

We show the 90\% C.L. upper limits on the annihilation cross section 
and the future HK sensitivity in the Fig.~\ref{fig;indirect}. 
For the $\psi\bar{\psi} \to \nu\bar{\nu}$ annihilation, 
we illustrate the similar limits derived in the previous works~\cite{Campo:2018dfh,Klop:2018ltd} for comparison. 
Our SK limit is consistent with their results below $m_\psi=50$\,MeV, 
while it is a little aggressive for higher mass. 
We guess the disagreement originates from the difference in modeling the $\nu_e\Ox$ cross section. 
We just assume the mono-energetic electron in the reaction, 
but in general the electron has a broad energy distribution. 
This will weaken the signal strength and decrease the sensitivity. 
For the secluded annihilation, the limits depend on the mass degeneracy.  
When $\psi$ and $X$ are highly degenerate ($m_X/m_\psi=0.99$), 
the spectrum has a very narrow width and looks the line shape within the detector resolution, 
so that the structure of the exclusion curve resembles the monochromatic one.  
On the other hand, the curve is stretched, which reflects the spectral shape that has the center at $E_\nu=m_\psi/2$. 
As $\psi$ and $X$ are less degenerate, the curve is shifting to the lighter mass. 
It is interesting that the exclusion curves extend to the high mass region above 300\,MeV due to the wide spectrum. 
The sensitivity in this mass region is rapidly decreasing 
as the lower energy cut of the neutrino flux, $E_- = y_- m_\psi$, exceeds about 100\,MeV. 
For instance, $E_- \simeq 0.28\,m_\psi$ with $m_X/m_\psi=0.9$, 
so that $m_\psi \sim 350$\,MeV is the boundary. 
In the high mass region, the sensitivity may be increased by combining other observations, 
such as measurements of the atmospheric neutrino flux by the SK~\cite{Richard:2015aua}. 

\begin{figure}[t]
\includegraphics[width=0.5\textwidth]{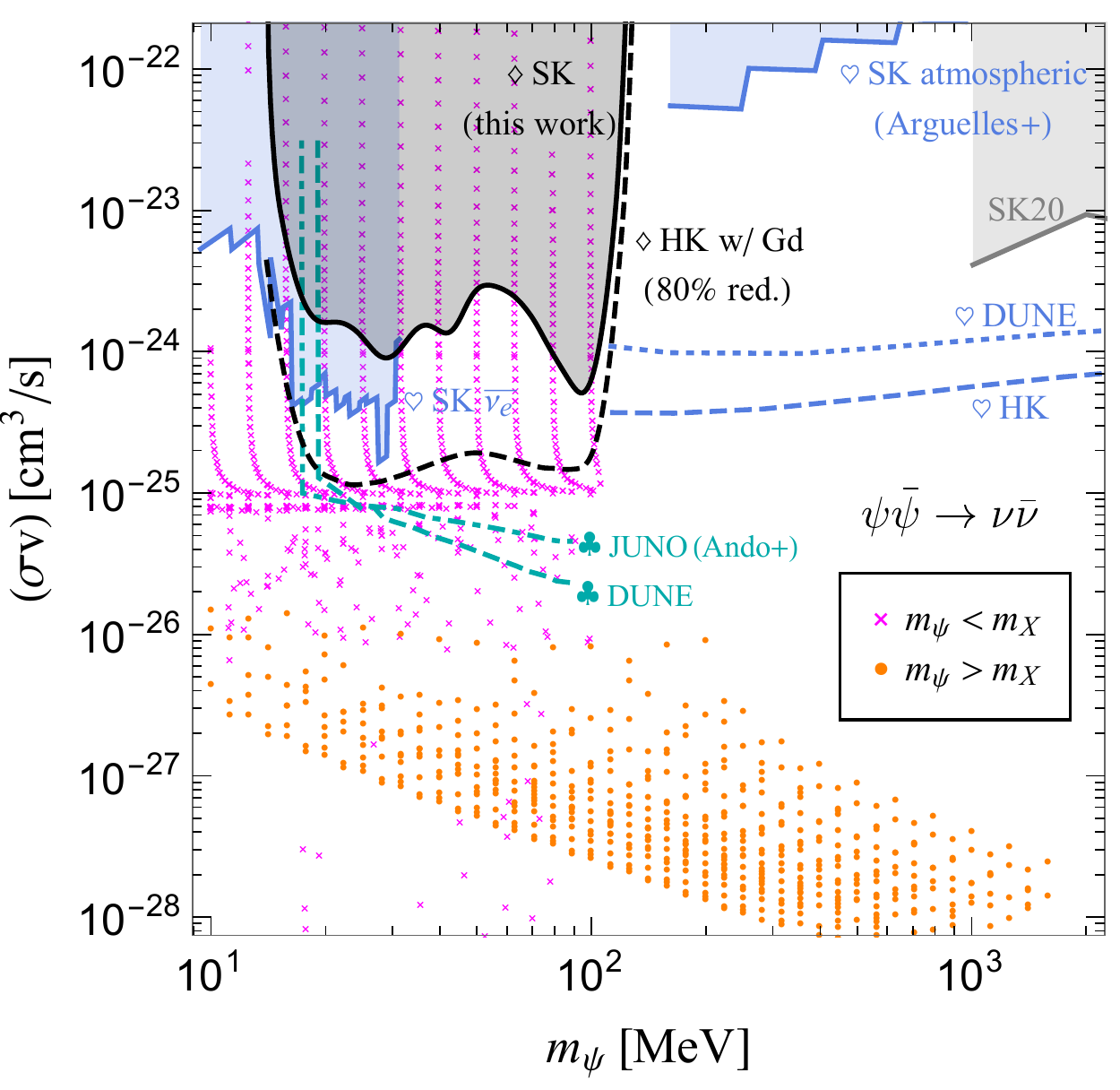}
\includegraphics[width=0.5\textwidth]{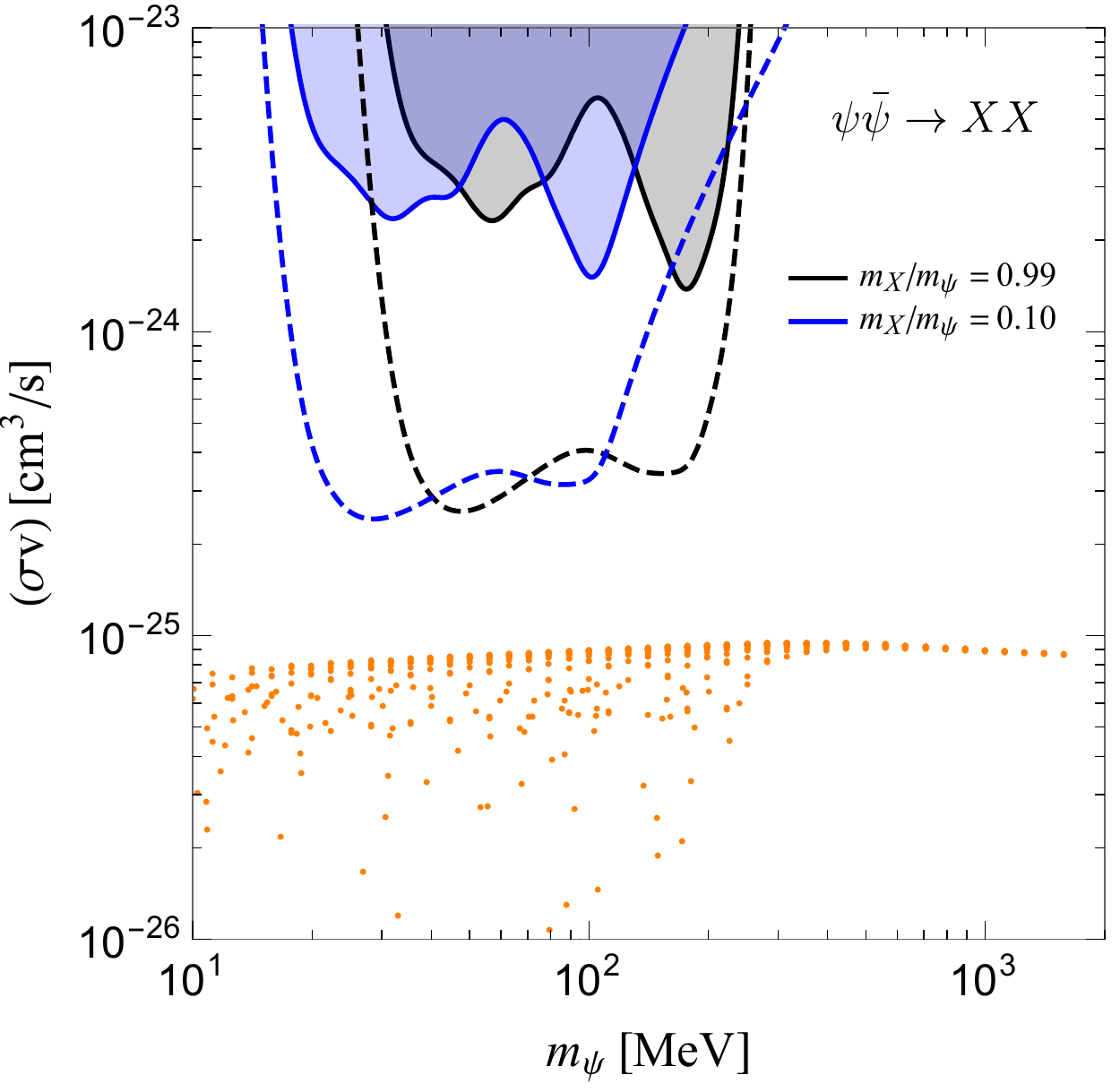}
\caption{The predicted annihilation cross section for $\psi\bar{\psi}\to\nu\bar{\nu}$ (left) and $\psi\bar{\psi} \to XX$ (right) cases. 
The magenta and orange points correspond to $m_\psi < m_X$ and $m_\psi > m_X$, respectively. 
All these points can explain the observed DM abundance and the muon $g-2$ as well as being consistent with the experimental constraints in Fig.\ref{fig;allowed}. 
The various sensitivity curves presented in the literature are shown, 
in addition to our results (black lines with the $\diamondsuit$ mark). 
The lines with the same mark ($\clubsuit$, $\heartsuit$) are extracted from the same paper:  
$\clubsuit$ Ando et al. \cite{Klop:2018ltd}, $\heartsuit$ Arguelles et al. \cite{Arguelles:2019ouk} and SK20~\cite{Abe:2020sbr}. 
To avoid being messy, we omit the similar lines shown in Fig.~\ref{fig;indirect}. 
}
\label{fig;indirect_mutau}
\end{figure}

In both annihilation modes, 
the current SK limits are much above the canonical thermal relic cross section, 
but the future HK sensitivity may reach down to it. 
We would like to add that Ref.~\cite{Klop:2018ltd} also studies the reach of 
DUNE~\cite{Acciarri:2015uup} and JUNO~\cite{An:2015jdp} experiments, 
that is more sensitive than the HK in the same mass range. 
These sensitivity plots are shown in Fig.~\ref{fig;indirect_mutau} (left). 
We also note that the difference in modeling the $\nu_e \Ox$ cross section will affect the exclusion limits in the secluded annihilation as well. 
Nevertheless, we expect that the modeling error is not so large compared with the monochromatic case unless DM and $X$ are highly degenerate. 
This is mainly because the neutrino flux is basically broad in the secluded annihilation, in contrast to the monochromatic one, and hence the electron event shape is also broad without the electron energy distribution in the $\nu_e \Ox$ reaction. 
Thus, the exclusion limits will not largely be changed even if another modeling generates a broad electron distribution in the reaction. 
The detailed study of modeling the $\nu_e \Ox$ cross section and its potential error is beyond the scope of this paper and will be followed in a future work.

It may be useful to comment on how large impact the polarizations of the intermediate state have on the cross section limits in Fig.~\ref{fig;indirect} (right). 
If the intermediate state in the secluded annihilation is an unpolarized vector or a scalar, the neutrino flux is box-shape. 
We have analyzed the neutrino signature and calculated the upper limit on the cross section for such a neutrino spectrum. 
Comparing the results with those of the bowl-shape spectrum, we have found the difference is at most 10--20\%. 
Since in general the neutrino spectrum is a medium of the box and bowl-shape, 
we expect the cross section limit to be comparable with the ones in Fig.~\ref{fig;indirect} (right), 
independently of the annihilation fraction into each polarization state.

\subsection{Implication for U(1)$_{\mutau}$ DM}

In the above subsections, we have formulated the analysis of neutrino signature 
and evaluated the experimental reach to the extra neutrino flux from the DM annihilation 
with neutrino telescopes. 
In this subsection, we discuss the impact on the U(1)$_{\mutau}$ DM. 

In Fig.~\ref{fig;indirect_mutau}, 
we plot the predictions of the annihilation cross section for the parameter points 
that can explain the observed DM abundance and the muon $g-2$ 
as well as being consistent with the experimental constraints in Fig.~\ref{fig;allowed}. 
The magenta and orange points give the cross sections for $m_\psi < m_X$ and $m_\psi > m_X$, respectively. 
The velocity of DM is assumed to be a typical virial velocity in the galaxy, $v_{\rm rel} \sim 10^{-3}$. 
We also summarize various limits on the cross section derived in the literature, 
in addition to our results (black lines with the $\diamondsuit$ mark). 
To avoid the messy figure, however, we omit the similar limits that are already shown in Fig.~\ref{fig;indirect}. 
The lines with the same mark ($\clubsuit$, $\heartsuit$) are extracted from the same paper:  
$\clubsuit$ Ando et al. \cite{Klop:2018ltd}, $\heartsuit$ Arguelles et al. \cite{Arguelles:2019ouk} and SK20~\cite{Abe:2020sbr}. 
In referring to these limits, we appropriately adjust the neutrino flavor fraction to $\kappa=1/5$. 

We see in Fig.~\ref{fig;indirect_mutau} (left) that 
for $m_\psi < m_X$, the model predicts so large cross section that the future experiments can reach. 
For some parameter points, 
the cross section is significantly boosted to have even 
$\VEV{\sigma v} = 10^{-22}\,{\rm cm^3/s}$. 
Such parameter points correspond to the resonant mass region $m_\psi \simeq m_X/2$. 
If the DM mass is fine-tuned to such values at or below \% level, 
the thermal history of the DM production is modified to delay the freeze-out. 
Such a delay requires the larger cross section than the standard case 
to deplete the number density down to the observed value. 
The mechanism for boosting the cross section in this way is discussed in \cite{Ibe:2008ye,Ibe:2009dx}. 
We review the detail of this mechanism in the Appendix. 
It is interesting that some of such resonant parameter sets have already been excluded by the SK data. 
When the mass tuning to $m_\psi \simeq m_X/2$ is moderate, on the other hand, 
suppressed cross sections can also be obtained. 
These are distributed over $\VEV{\sigma v} \lesssim {\cal O}(10^{-26})\,{\rm cm^3/s}$ in the figure (magenta points). 
The suppression is caused if the cross section is highly enhanced at the typical freeze-out time 
due to the physical $X$ boson resonance. 
In this case, the small DM charge is enough to thermally produce the DM abundance. 
As a result, the late-time annihilation, e.g. in the galaxy, is suppressed 
because the DM velocity in the galaxy is too small to produce the $X$ boson resonance. 
This region corresponds mainly to $q_\psi = {\cal O}(1)$. 
In the non-resonant region, the DM thermal production works in the standard way, 
so that we have an almost constant canonical cross section, $\VEV{\sigma v} \simeq 10^{-25}\,{\rm cm^3/s}$. 
None of the current experiments can reach the canonical value, 
but the future experiments including DUNE and JUNO will be able to cover the DM mass of 20--100\,MeV.

In the secluded region ($m_\psi > m_X$), the direct annihilation into a neutrino pair is suppressed 
and the size of the cross section is at most $\sim10^{-26}\,{\rm cm^3/s}$. 
This tendency is more pronounced as DM is heavier. 
In this case, however, the secluded annihilation can be large (see Fig.~\ref{fig;indirect_mutau} (right)). 
We see the model predictions distributed slightly below the canonical value, 
$\VEV{\sigma v} = 10^{-25}\,{\rm cm^3/s}$. There is no enhancement in this mass region. 
The cross section can be kinematically suppressed in the mass degenerate case, 
because the phase-space of the produced $X$ boson becomes small for the low DM velocity. 
Thus, a moderate mass splitting is favored to observe the signal, 
although the high degeneracy produces the sharp neutrino spectrum and leads the strong limits. 
Indeed, we see some parameter points lying much below the canonical value, 
where DM and the $X$ boson have close mass. 
The current SK limit is over one order of magnitude larger than the predicted cross section. 
The estimate of the future HK reach is above the prediction by a factor of 3 or more. 
We hope that some future updates or improvements of the analysis will grow 
the experimental sensitivity and reach the thermal relic cross section. 
Compared the HK sensitivity with the DUNE and JUNO ones in Fig.~\ref{fig;indirect_mutau} (left), 
the latter ones have better sensitivity. 
If this is the case for the secluded annihilation, 
it will be important to analyze the similar signals at these neutrino observatories, 
that will be pursued in future. 

It is worth mentioning other possibilities that 
predict significant neutrino flux from sub-GeV DM. 
Indeed, a great deal of effort has been devoted to non-renormalizable DM-neutrino interactions. 
The prime difficulty in embedding them in a renormalizable model is 
that the model includes interactions with charged leptons, 
which are isospin partners of neutrinos. 
Such interactions make DM or a mediator particle visible in terrestrial experiments and cosmological observations and, hence, will strongly limit the categories that achieve the sizable DM-neutrino interactions in a renormalizable manner. 
The model considered here is one of the simplest renormalizable models, 
that evades the experimental constraints and 
does not suffer from theoretical requirements, such as gauge anomalies. 
The other type of possibility is realized by employing a $t$-channel mediator. 
Examples of feasible renormalizable models incorporating a $t$-channel mediator include \cite{Batell:2017cmf,McKeen:2018pbb,Blennow:2019fhy,Okawa:2020jea}.

\section{Summary and discussions}
\label{sec:summary}

We have examined a simple Dirac DM model based on the U(1)$_{\mutau}$ gauge theory. 
In this model, DM $\psi$ interacts with the SM particles only through the $X$ boson. 
DM is produced thermally in the early Universe and annihilates into the SM particles through 
two kinds of processes. 
One is the $s$-channel $\psi \bar{\psi} \to f \bar{f}$ which is important for $m_{X}>m_{\psi}$. 
The other is the $t$-channel $\psi \bar{\psi} \to X X$. 
This annihilation process is allowed if $m_{X}<m_{\psi}$, and becomes significant for 
a large U(1)$_{\mutau}$ charge $q_{\psi}$ of DM.  
In both cases, the annihilation cross sections can be large enough even at off-resonance 
of $X$ boson mass pole. 
We have shown that the observed relic abundance of DM and the discrepancy in the 
muon anomalous magnetic moment are explained simultaneously by 
the U(1)$_{\mutau}$ gauge boson $X$ without conflicting the severe experimental bounds 
thanks to no interaction with SM quarks and electron.  
As a by-product of a large $q_{\psi}$ scenario, 
DM annihilations provide characteristic neutrino signatures at SK and HK.

In this paper, we have formulated the analysis of the indirect neutrino signal of DM 
in a model independent way, and applied to U(1)$_{\mutau}$ DM model. 
From the $\psi \bar{\psi} \to \nu \bar{\nu}$ process, 
neutrinos with monochromatic energy are predicted, 
while from the $\psi \bar{\psi} \to X X$ process followed by $X\to\nu\bar\nu$, 
the energy spectrum of neutrinos is bowl-shape. 
Neutrinos produced in the DM halo can be detected by neutrino telescopes. 
We have calculated the number of the signal events expected at SK and HK detectors, 
and estimated the upper limits from the future HK sensitivity to the DM annihilation cross section.
As we have shown in Fig.~\ref{fig;indirect_mutau}, 
the future sensitivity to the annihilation cross section at HK almost reaches the canonical thermal relic 
cross section in the $\psi \bar{\psi} \to \nu \bar{\nu}$ case. 
On the other hand, in the $\psi \bar{\psi} \to X X \to 2 \nu 2 \bar{\nu}$ case, 
the obtained future sensitivity is several times larger than the canonical one. 
Further improvement of the experimental sensitivity is necessary 
in order to cover the wide range of the model predictions. 
We hope that the analysis with directional information may help the background subtraction, 
which is beyond the scope of this paper. 
It will be shown elsewhere.

\section*{Acknowledgement}

We thank Julian Heeck for a valuable comment on an effect of vector boson polarizations on the neutrino spectral shape in the secluded annihilation. 
KA and SO would like to thank the hospitality of the Particle Theory Group of Kyushu University, where this work was initiated during their visit. 
This work is supported in part by JSPS KAKENHI Grant Numbers JP19J13812(KA) 
and JP18H05543(KT) and NSERC of Canada (SO). 
The authors thank the Yukawa Institute for Theoretical Physics at Kyoto University. Discussions during the YITP workshop YITP-W-20-08 on “Progress in Particle Physics 2020” were useful to complete this work. 

\appendix

\section{Breit-Wigner enhancement at indirect detection}

As shown in Fig.~\ref{fig;indirect_mutau}, 
there is a significant enhancement in indirect detection via the unphysical $X$ boson pole. 
In this appendix, we briefly review the effect. 
We consider the $\psi\bar{\psi} \to \nu\bar{\nu}$ process. 
Assuming the initial state DM is non-relativistic, we express the center of mass energy and DM mass as 
\begin{equation}
s \simeq 4 m_\psi^2 + m_\psi^2 v_{\rm rel}^2 ,\quad 
4 m_\psi^2 = m_X^2 (1+\delta) ,
\end{equation}
where $v_{\rm rel}$ denotes the relative velocity of DM. 
Then, the annihilation cross section is written by
\begin{equation}
(\sigma v)_{\nu\bar{\nu}} \simeq \frac{q_\psi^2 g_X^4}{\pi} \frac{1}{m_X^4} \frac{m_\psi^2}{(\delta+v_{\rm rel}^2/4)^2+\gamma_X^2} ,
\end{equation}
where $\gamma_X = \Gamma_X/m_X$. 

Let us consider how the DM freeze-out process is changed for $\delta \ll 1$. 
First of all, we approximate the cross section 
\begin{equation}
(\sigma v)_{\nu\nu} \approx 
\frac{q_\psi^2 g_X^4}{\pi} \frac{m_\psi^2}{m_X^4} \times 
\left\{ \begin{matrix} 
({\rm Max}[v_{\rm rel}^2/4,\gamma_X])^{-2} & (v_{\rm rel}^2 > \delta) \\
({\rm Max}[\delta,\gamma_X])^{-2} & (v_{\rm rel}^2 < \delta)
\end{matrix} \right.
\end{equation}
It follows from this equation that as DM velocity decreases with the expanding universe, 
the cross section becomes large as $(\sigma v)_{\nu\bar{\nu}} \propto v_{\rm rel}^{-4}$. 
The increase of the cross section continues until 
$v_{\rm rel}^2 \lesssim {\rm Max}[\delta, \gamma_X]$ is satisfied. 
In this case, the DM number density does not freeze out at the typical freeze-out temperature 
$x_f =m_\psi/T_f \sim 20$ ($v_{\rm rel}^2 \sim 0.1$), and 
continues to decrease via the annihilation even at lower temperature.
As a result, the actual freeze-out time is delayed and the produced thermal abundance is modified to be 
\begin{equation}
\Omega h^2 \sim 0.1 \times \left(\frac{10^{-26}\,{\rm cm^3/s}}{\VEV{\sigma v}_{T=0}}\right) \frac{x_b}{x_f} 
\label{eq:Omega_BF} ,
\end{equation}
where $x_f \simeq 20$ is the typical freeze-out temperature and $x_b$ is the actual one. 
The actual freeze-out temperature is given by~\cite{Ibe:2008ye,Ibe:2009dx}
\begin{equation}
\frac{1}{x_b} \simeq \frac{1}{\VEV{\sigma v}_{T=0}} \int_{x_f}^\infty \frac{\VEV{\sigma v}}{x^2} dx 
\simeq {\rm Max}[\delta, \gamma_X] .
\end{equation}
It suggests that the cross section is boosted by a factor of $x_b/x_f$. 
${\rm BF}=x_b/x_f$ is called the boost factor. 
We have $\gamma_X \propto g_X^2 \lesssim 10^{-6}$ in the model, 
the boost factor can be as large as $10^6$ in a case. 

Indirect detection experiments observe the neutrinos from DM annihilation in our Galaxy. 
Since the DM velocity is $v_{\rm rel} \sim 10^{-3}$ at most, 
we can approximate $(\sigma v)_{\nu\nu} \simeq \VEV{\sigma v}_{T=0}$ for e.g. $\delta=10^{-3}$. 
In this case, the boost factor is expected to have 
\begin{equation}
{\rm BF} \simeq \frac{1}{10\,{\rm Max}[\delta,\gamma_X]} \sim 100
\end{equation}
leading to the enhanced cross section, 
\begin{equation}
(\sigma v)_{\nu\bar{\nu}} \sim {\rm BF} \times 10^{-26}\,{\rm cm^3/s} .
\end{equation}

Before closing, we derive Eq.(\ref{eq:Omega_BF}). 
In the standard way, the approximate solution of the Boltzmann equation is obtained 
by integrating the equation, 
\begin{equation}
\frac{dY_{\rm DM}}{dx} \simeq - \frac{\lambda}{x^2} Y_{\rm DM}^2 ,
\end{equation}
over $[x_f,\infty]$, where 
$x_f$ is determined by appropriately matching the approximate solution with the actual one. 
Above, we assumed the $s$-wave DM annihilation and 
introduced $Y_{\rm DM}=n_{\rm DM}/s$ with $s$ being the entropy density and 
\begin{equation}
\lambda = \sqrt{\frac{8\pi^2}{45} g_*} M_{\rm pl} m_{DM} \VEV{\sigma v}_{T=0} .
\end{equation}
Taking into account the resonant behavior of the cross section, the equation is modified as 
\begin{equation}
\frac{dY_{\rm DM}}{dx} \simeq - \frac{\lambda}{x^2} \frac{\VEV{\sigma v}_T}{\VEV{\sigma v}_{T=0}} Y_{\rm DM}^2 .
\end{equation}
One can readily get Eq.(\ref{eq:Omega_BF}) by solving this equation.


{\small
\bibliographystyle{JHEP}
\bibliography{ref_MuTauDM}
}

\end{document}